\documentclass[onecolumn]{emulateapj}

\usepackage{graphicx}

\newcommand {\omegape}	{\omega_\mathrm{pe}}
\newcommand {\nez}	{n_\mathrm{e0}}
\newcommand {\me}	{m_\mathrm{e}}
\newcommand {\mi}	{m_\mathrm{i}}
\newcommand {\Vsh}	{V_\mathrm{sh}}
\newcommand {\MA} {M_\mathrm{A}}
\newcommand {\vA} {v_\mathrm{A}}
\newcommand {\lambdae}	{\lambda_\mathrm{e}}
\newcommand {\lambdai}	{\lambda_\mathrm{i}}
\newcommand {\omegapi}	{\omega_\mathrm{pi}}
\newcommand {\omegace}	{\omega_\mathrm{ce}}
\newcommand {\omegaci}	{\omega_\mathrm{ci}}
\newcommand {\kDe} {k_\mathrm{De}}
\newcommand {\kDi} {k_\mathrm{Di}}

\newcommand {\kB}	{k_\mathrm{B}}
\newcommand {\Te}	{T_\mathrm{e}}
\newcommand {\Ti}		{T_\mathrm{i}}
\newcommand {\Ekin}	{E_\mathrm{kin}}

\newcommand {\omegaps}	{\omega_\mathrm{ps}}
\newcommand {\omegacs}	{\omega_\mathrm{cs}}
\newcommand {\zetas}  {\zeta_\mathrm{s}}
\newcommand {\kDs}    {k_\mathrm{Ds}}
\newcommand {\ns}     {n_\mathrm{s}}
\newcommand {\Vs}     {V_\mathrm{s}}
\newcommand {\Vsx}     {V_{\mathrm{s},x}}
\newcommand {\Vsz}     {V_{\mathrm{s},z}}
\newcommand {\Ts}     {T_\mathrm{s}}
\newcommand {\as}     {a_\mathrm{s}}
\newcommand {\ms}     {m_\mathrm{s}}

\newcommand {\alps}     {\alpha_\mathrm{s}}
\newcommand {\tVsx}     {\tilde{V}_{\mathrm{s}x}}
\newcommand {\tVsz}     {\tilde{V}_{\mathrm{s}z}}
\newcommand {\tVs}     {\tilde{V}_{\mathrm{s}}}
\newcommand {\etas}     {\eta_\mathrm{s}}

\shorttitle{WEAKLY MAGNETIZED COLLISIONLESS SHOCKS}
\shortauthors{KATO AND TAKABE}

\begin{document}

\title{Nonrelativistic collisionless shocks in weakly magnetized electron--ion plasmas:
two-dimensional particle-in-cell simulation of perpendicular shock}

\author{Tsunehiko N. Kato and Hideaki Takabe}
\affil{Institute of Laser Engineering, Osaka University,
2-6 Yamada-oka, Suita, Osaka 565-0871, Japan}
\email{kato-t@ile.osaka-u.ac.jp}

\begin{abstract}
A two-dimensional particle-in-cell simulation is performed
to investigate weakly magnetized perpendicular shocks
with a magnetization parameter of $\sigma = 6 \times 10^{-5}$,
which is equivalent to a high Alfv\'en Mach number $\MA$ of $\sim 130$.
It is shown that current filaments form in the foot region of the shock
due to the ion-beam--Weibel instability
(or the ion filamentation instability)
and that they generate a strong magnetic field there.
In the downstream region,
these current filaments also generate a tangled magnetic field that is
typically 15 times stronger than the upstream magnetic field.
The thermal energies of electrons and ions in the downstream region
are not in equipartition and their temperature ratio is $\Te/\Ti \sim 0.3 - 0.4$.
Efficient electron acceleration was not observed in our simulation,
although a fraction of the ions are accelerated slightly on reflection
at the shock.
The simulation results agree very well with the Rankine--Hugoniot relations.
It is also shown that electrons and ions are heated in the foot region
by the Buneman instability (for electrons)
and the ion-acoustic instability (for both electrons and ions).
However, the growth rate of the Buneman instability is significantly reduced due to
the relatively high temperature of the reflected ions. For the same reason,
ion--ion streaming instability does not grow in the foot region.
\end{abstract}

\keywords{instabilities --- magnetic fields --- plasmas --- shock waves --- supernova remnants}

\section{Introduction}
A large volume of the universe
(including interstellar and intergalactic space)
is filled with hot, tenuous plasmas.
Coulomb collisions between charged particles
rarely occur in these plasmas and the plasma dynamics are dominated by
collective phenomena involving particles and electromagnetic fields
(e.g., plasma oscillations).
Hence, such plasmas are known as collisionless plasmas.
Even in collisionless plasmas,
some kinds of ``shocks'' occur.
These shocks generally have very complex formation mechanisms that involve various kinetic processes, including
electrostatic instabilities, electromagnetic instabilities, and compression of background magnetic fields.
The shocks driven in supernova remnants (SNRs) are
considered to be collisionless ones.

SNR shocks propagate in
the interstellar medium, which has weak magnetic fields of typically $\sim 3 \mu$G.
In the context of shocks in magnetized plasmas,
the strength of the magnetic field, $B_0$, is frequently expressed in terms of
the magnetization parameter or the sigma parameter, which is defined as the ratio of
the magnetic energy density to the bulk kinetic energy density of the upstream plasma
(both are measured in the shock rest frame).
For nonrelativistic cases,
it is given by
\begin{equation}
	\sigma \equiv \frac{B_0^2/8\pi}{\nez(\me + \mi)\Vsh^2/2} = \MA^{-2},
\end{equation}
where $\nez$ is the electron number density in the upstream plasma,
$\me$ is the electron mass, $\mi$ is the ion mass,
$\Vsh$ is the shock speed, and $\MA$ is the Alfv\'{e}n Mach number.
For example,
for the shock in SN1006 (except the North-West region),
$\nez \sim 0.05$ cm$^{-3}$ and $\Vsh \sim 4900$ km s$^{-1}$ 
were inferred \citep{Acero07}
so that $\sigma \sim 4 \times 10^{-5}$.
The recently discovered `youngest' SNR G1.9+0.3 is considered to have
a shock velocity of $\Vsh \sim 14,000$ km s$^{-1}$ \citep{Reynolds08};
assuming that $\nez \sim 0.1$ cm$^{-3}$ gives $\sigma \sim 2 \times 10^{-6}$.
The sigma generally lies in the range $10^{-6} < \sigma < 10^{-3}$
for shocks in young SNRs;
these shocks are thus very low-$\sigma$ shocks
(or, equivalently, very high Alfv\'en Mach number shocks).

Magnetized shocks have been extensively investigated,
especially perpendicular shocks
in which the background magnetic field is perpendicular to the shock normal.
The structure of perpendicular shocks in the supercritical regime
($\MA > \MA^*$, where $\MA^* \sim 3$)
is known to some extent:
a fraction of the incoming ions are reflected at the shock front (called the `ramp') and the reflected ions form a slightly dense
region, referred to as the `foot', in front of the ramp.
The ions also accumulate immediately behind the ramp
and generate a strong magnetic field there,
which is called the (magnetic) `overshoot'. 
Over the last decade,
one-dimensional (1D) particle-in-cell (PIC) simulations
that can model the kinetic dynamics of both electrons and ions
have been performed to investigate high-Mach-number perpendicular shocks
\citep[e.g.,][]{Shimada00, Schmitz02, Scholer03}.
Recently,
several two-dimensional (2D) simulations have also been performed
\citep[e.g.,][]{Umeda08, Amano09, Lembege09}.
However,
most simulations have been conducted for relatively strong background magnetic fields
($\sigma > 5 \times 10^{-3}$ or $\MA < 15$). It is thus desirable to perform
simulations for weaker background fields.

On the other hand,
it was recently demonstrated that
certain kinds of collisionless shocks can occur
even in unmagnetized plasmas
at relativistic shock speeds
by performing two- or three-dimensional (3D) PIC simulations
\citep{Kato07, Spitkovsky08, Chang08}.
In these shocks,
the beam--Weibel instability (or filamentation instability) is driven
in the transition region of the shocks
between the counterstreaming electron--positron beams in pair plasmas
or between the counterstreaming ion beams in electron--ion plasmas 
and generates strong magnetic fields there.
These generated fields provide an effective dissipation mechanism
for collisionless shock formation and are hence
often referred to as ``Weibel-mediated shocks.''
[Note that the beam--Weibel instability is driven by the counterstreaming
beams \citep[c.f.][]{Fried59} and it differs from the ordinary Weibel
instability, which is driven by a temperature anisotropy \citep{Weibel59} 
(see also \citet{Davidson72}).]
These shocks can be driven by relativistic phenomena, such as
gamma-ray bursts and their afterglows \citep{Medvedev99,Brainerd00}, jets from active galactic nuclei,
and pulsar winds \citep{Kazimura98}.
It was also shown in our previous paper \citep{Kato08}
that this kind of shock can form in unmagnetized electron--ion plasmas
even at nonrelativistic speeds.
The beam--Weibel instability can also be important in weakly magnetized shocks.
As was shown in our previous paper,
the magnetic field generated by the ion beam--Weibel instability
reaches a few percent of the upstream bulk kinetic energy
and this value is much higher than the background magnetic field around typical SNR shocks.
Therefore,
the ion beam--Weibel instability may play an important role in the formation
of weakly magnetized shocks.
Magnetized shocks have been extensively investigated by 1D simulations.
However,
1D simulations cannot consider the beam--Weibel instability 
because its wave vector is
perpendicular to the flow direction.
Therefore,
it is essential to perform multidimensional simulations to investigate the formation process of
weakly magnetized nonrelativistic shocks.

Collisionless shocks are also considered sites of particle acceleration.
In particular,
cosmic-rays with energies below $10^{15}$ eV are considered to be
accelerated in SNR shocks.
Indeed,
recent X-ray observations revealed that
electrons are accelerated to energies of $\sim 10^{14}$ eV
in several young SNRs \citep{Koyama95, Long03, Bamba03}.
It is widely accepted that
first-order Fermi acceleration
or diffusive shock acceleration is the acceleration mechanism 
\citep[e.g.,][]{Drury83, Blandford87}.
However,
it is currently not possible to determine 
the fraction of thermal plasma particles that are injected
into the diffusive shock acceleration process (this is known as the injection problem).
For electron injection in quasi-perpendicular shocks,
the shock surfing acceleration
has been investigated as an injection mechanism
or even as an efficient acceleration mechanism
\citep{Hoshino01, McClements01, Hoshino02}.
However,
several researchers have recently shown
that the shock surfing acceleration process is in fact inefficient
in 2D \citep[e.g.,][]{Dieckmann06, Ohira07, Umeda08} and that
it is efficient in 1D because of the symmetry of the system.
Instead,
\citet{Amano09} showed that
another acceleration process can operate in 2D 
in which a fraction of electrons are reflected in the foot region
by small-scale electrostatic waves generated by the Buneman instability.
They are then accelerated by the motional electric field as well as
being directly accelerated by the electric field when they resonate with the electrostatic waves.
Electrons can be accelerated up to
about the upstream ion bulk energy by this mechanism.
Thus,
multidimensional effects can play an essential role in
the acceleration/injection mechanism.

In addition,
collisionless shocks can be sites of magnetic field amplification/generation.
Recent X-ray observations suggest that magnetic fields of the order of hundreds of microgauss or even milligauss may be
generated in the vicinity of SNR shocks
\citep{Vink03, Voelk05, Uchiyama07}.
Several mechanisms have been proposed for this magnetic field amplification, including
a nonresonant instability driven by high-energy particles accelerated
in shocks \citep{Bell04} and 
magnetohydrodynamic turbulence behind shocks \citep{Giacalone07, Inoue09}.
The mechanism may be related with the microscopic kinetic processes
associated with shock formation itself;
it should in principle be possible to investigate this by performing large-scale PIC simulations.

In this study,
we investigate the formation and structure of perpendicular shocks
for very low $\sigma$
and the processes responsible for particle acceleration and magnetic field generation
by performing 2D PIC simulation,
which can appropriately model the beam--Weibel instability.
Because of the capability of the computer,
we used a reduced ion-to-electron mass ratio
and a shock speed ($\Vsh \sim 0.3c$, where $c$ is the speed of light) that is much higher than realistic ones
for SNRs ($\Vsh \sim 0.01c$) in the simulation.

\section{METHOD}
We investigated collisionless shocks
in electron--ion plasmas with weak background magnetic fields
by performing a 2D PIC simulation.
The simulation code is a relativistic, electromagnetic, PIC code
with two spatial and three velocity dimensions
developed based on a standard method described by \citet{Birdsall}.
The basic equations of the simulation are
Maxwell's equations
and the (relativistic) equation of motion for particles.
In the following,
we regard the simulation plane as the $x-y$ plane
and we take the $z$-axis to be perpendicular to the plane.
We take $\tau = \omegape^{-1}$ to be the unit of time
and the electron skin depth $\lambdae = c \omegape^{-1}$ to be the unit of length,
where $\omegape \equiv (4\pi \nez e^2 / \me)^{1/2}$
is the electron plasma frequency defined for the electron
number density in the far upstream region, $\nez$.
The units for electric and magnetic fields are $E_* = B_* = c (4\pi \nez \me)^{1/2}$.

In the simulation,
a collisionless shock is driven
according to the ``injection method.''
There are two walls, one on the left-hand side (smaller $x$)
and the other on the right-hand side (larger $x$) of the simulation box
and these walls reflect particles specularly.
Initially, both electrons and ions are loaded uniformly
in the region between the two walls
with a bulk velocity of $V$ in the $+x$-direction.
The electrons and ions have equal temperatures
in the upstream region.
In the early stages of the simulation,
particles near the right wall
are reflected by the wall
and then interact with incoming particles
(i.e., the upstream plasma).
This interaction generates some instability
and eventually a collisionless shock forms.
The frame of the simulation
is the rest frame of the shock downstream;
the shock propagates
from right to left in the downstream rest frame.

We consider a perpendicular shock in this paper;
the initial magnetic field, $B_0$, is in the $y$-direction
(i.e., in the plane)
and its strength is determined by the sigma.
However,
since the shock speed is unknown before performing the simulation,
in the following, the sigma is defined in the simulation frame
with an upstream bulk velocity $V$ instead of the shock speed $\Vsh$ as
\begin{equation}
	\tilde{\sigma} \equiv \frac{B_0^2/8\pi}{\nez(\me + \mi)V^2/2};
\end{equation}
however, the difference between these two sigmas is not large.
With this definition of the sigma,
the magnetic field strength in the simulation frame is given by
$B_0 = \left[ (1 + \mi/\me ) \tilde{\sigma} \right]^{1/2} (V/c) B_*$.
The initial electric field, $E_0$, is determined so that it vanishes
in the plasma rest frame (i.e., the upstream frame);
this requirement causes the motional electric field in the simulation frame,
$E_0 = -VB_0/c$, in the $z$-direction.
The boundary conditions for both the particles and the electromagnetic field
are periodic in the $y$-direction.

\section{RESULTS AND ANALYSIS}
We performed a simulation for a sigma of $\tilde{\sigma} = 10^{-4}$.
As mentioned above,
we use a reduced ion mass of $\mi = 30 \me$
and a bulk velocity of $V = 0.25c$.
The grid size is $N_x \times N_y = 16384 \times 1024$
and there are $\sim 40$ particles per cell per species.
The physical dimensions of the simulation box are
$L_x \times L_y = 3200\lambdae \times 200\lambdae$
and thus the size of a cell is $\Delta x = \Delta y \sim 0.2\lambdae$.
The electron and ion temperatures are equal and
are given by $\kB T/\me c^2 = 1.25 \times 10^{-3}$,
where $\kB$ is the Boltzmann constant.
The thermal velocities are thus given by
$a_\mathrm{e} = (2\kB T/\me)^{1/2} = 0.05c$ for the electrons
and $a_\mathrm{i} = 9.13 \times 10^{-3}c$ for the ions.
For these parameters, we have
$\omegace / \omegape = B_0/B_* \sim 1.4 \times 10^{-2}$,
the Alfv\'en speed $\vA \sim 2.5 \times 10^{-3}c$
(thus, $\tilde{\MA} \equiv V/\vA = 100$),
and the plasma beta $\beta \sim 26$ (i.e., it is a high-beta plasma).
The Larmor radii of the electrons and ions
calculated for the background field and the upstream bulk velocity
are $r_{g,e} = 18\lambdae$
and $r_{g,i} = 535\lambdae$, respectively.

\subsection{Overall structure}
\begin{figure}[htbp]
\plotone{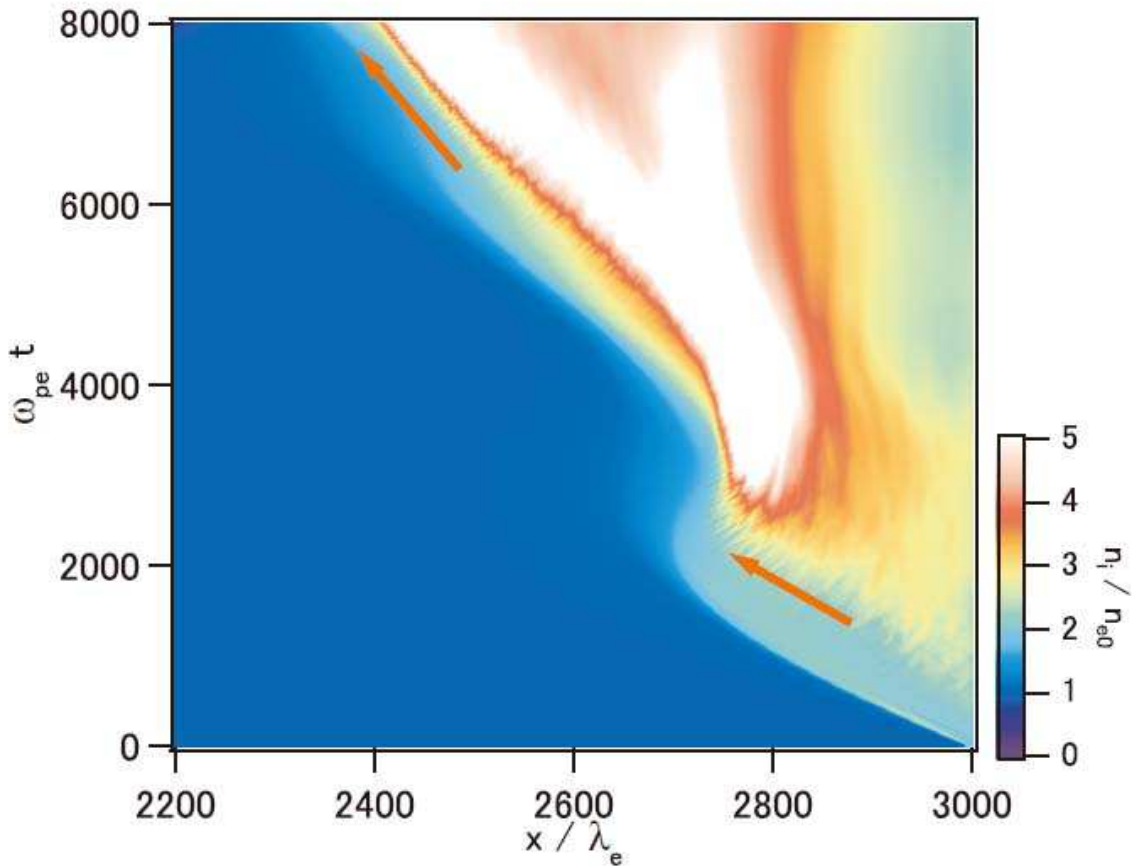}
\caption{
Time development of the ion number density averaged over the $y$-direction
for the simulation with an upstream bulk velocity of $V = 0.25c$ and
a sigma parameter of $\tilde{\sigma} = 10^{-4}$.
The color indicates the number density normalized by that in the far upstream $\nez$.
The horizontal and vertical axes respectively represent $x$ (in units of electron skin depth $\lambdae$)
and time (in units of the electron plasma time $\omegape^{-1}$).
The shock structure and its propagation speed change abruptly
around $\omegape t \sim 3000$
due to the transition from an unmagnetized shock to a magnetized shock.
The arrows indicate the obtained shock speeds for the unmagnetized shock ($V_\mathrm{sh,d} \sim -0.16c$)
and the magnetized shock ($V_\mathrm{sh,d} \sim -0.08c$).
}
\label{fig:ni_dev}
\end{figure}

Figure \ref{fig:ni_dev} shows the time evolution of the ion number density
averaged over the $y$-direction.
The shock transition region, or the ``shock front'', appears as a steep increase in the number density.
The shock structure and its propagation speed abruptly change around $\omegape t \sim 3000$.
This is because the shock structure undergoes a transition
from an unmagnetized shock to a magnetized shock.
Indeed, the structure for $\omegape t < 2000$ is
essentially the same as those of Weibel-mediated shocks in unmagnetized plasmas
\citep{Kato08}, as discussed below.
The transition time is of the order of the gyration time of the ions
in the background field, $T_g \equiv 2\pi/\omegaci$,
where $\omegaci \equiv e B_0/\mi c$ is the ion cyclotron frequency.
For the unmagnetized shock ($1500 < \omegape t < 2000$),
the shock speed measured in the downstream frame is
$V_\mathrm{sh,d} \sim -0.16c$.
For the magnetized shock ($\omegape t > 4000$),
it becomes $V_\mathrm{sh,d} \sim -0.08c$ and
that in the upstream frame and the Alfv\'en Mach number
are given by $V_\mathrm{sh} \sim -0.33c$
and $\MA \sim 130$, respectively.
Thus,
the sigma defined for the shock velocity
is given by $\sigma = 5.9\times 10^{-5}$ in this case.
The shock speeds obtained here may have small uncertainties
because they were obtained by eye-fitting the figure and also
they may not be in the steady state yet.
Since the formation of an unmagnetized shock is a consequence of the initial conditions
and we are interested in the magnetized shock in this study,
we mainly focus on the magnetized shock below.
We discuss the unmagnetized shock at the end of this section.

\begin{figure}[htbp]
\plotone{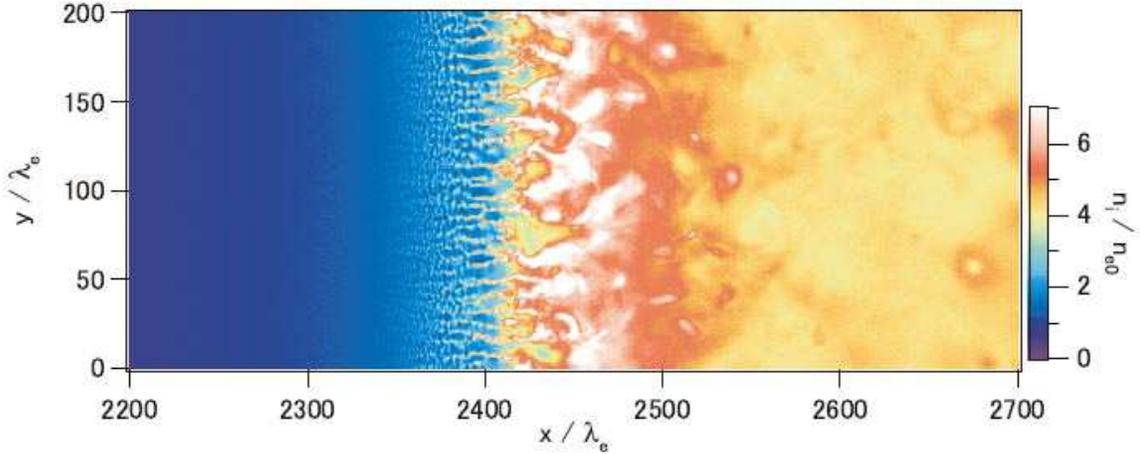}
\caption{
Ion number density at $\omegape t = 8000$.
The horizontal and vertical axes represent $x$ and $y$, respectively.
The left- and right-hand sides are upstream and downstream of
the shock, respectively. Many filamentary structures can be seen
in the shock transition region ($2350 < x / \lambdae < 2550$).
}
\label{fig:ni_160000}
\end{figure}

Figure~\ref{fig:ni_dev} shows that
the shock wave almost reaches steady state near the end of the simulation ($\omegape t \sim 8000$).
Figure \ref{fig:ni_160000} shows the ion number density at $\omegape t = 8000$.
(Hereafter, we discuss the results at this time unless otherwise stated.)
The upstream plasma flows from left to right
and moves through the transition region ($2350 < x/\lambdae < 2550$)
and then reaches the downstream state.
(The structure in Fig.~\ref{fig:ni_dev} in $x > 2700 \lambdae$
is an artifact due to the boundary
and so in the following we discuss the structure in $x < 2700 \lambdae$.)
Note that there are filamentary structures, which cannot be observed in 1D simulations, in the upstream leading edge of the shock transition region ($x \sim 2400 \lambdae$).
The filament radius is typically approximately equal to the ion inertial length,
which is the same as those in the ``Weibel-mediated'' shocks
in unmagnetized electron--ion plasmas \citep{Kato08}.
Then, behind them,
there is a highly fluctuating high-density region.
In the downstream region ($x > 2550 \lambdae$),
the number density becomes almost homogeneous.

\begin{figure}[htbp]
\plotone{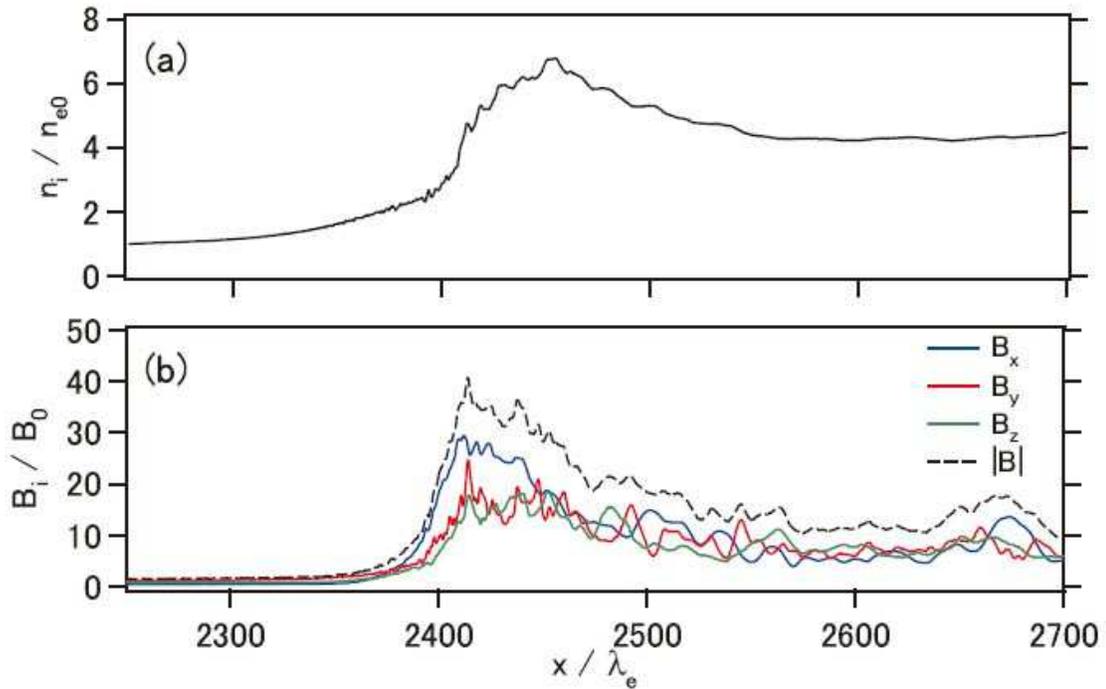}
\caption{
Profiles of (a) the ion number density normalized by the upstream density $\nez$
and (b) the root mean square of each magnetic field component
($B_x$, blue curve; $B_y$, red curve; $B_z$, green curve) 
and that of the total strength ($|B|$, dashed black curve),
where all the components are normalized by the upstream background field $B_0$.
}
\label{fig:profiles}
\end{figure}

Figures \ref{fig:profiles}(a) and (b) show profiles of the ion number density and the magnetic field strength averaged over the $y$-direction, respectively.
It shows that the number density increases rapidly
in the transition region and after the transition region 
it approaches $\sim 4$ times the upstream value.
Figure~\ref{fig:profiles}(b) shows
the root mean square of each magnetic field component
together with the total magnetic field strength.
This structure is similar to the well-known structure of 
supercritical perpendicular shocks in 1D;
the `ramp' is at $x \sim 2400 \lambdae$
and there is an extended `foot' region in $x < 2400 \lambdae$
as well as an `overshoot' region in $2400 < x/\lambdae < 2470$.
It is evident that
strong magnetic fields are generated in both the shock transition region
(or the overshoot)
and the downstream region.
The energy density of the magnetic field
reaches $\sim 15\%$ of the upstream bulk kinetic energy density (measured in the downstream rest frame)
in the shock transition region
and $\sim 2\%$ in the downstream region.
It is notable that
$B_x$ and $B_z$, which are generated by the current filaments of the ion-beam--Weibel instability (see below),
are comparable with $B_y$, which is mostly generated by the upstream background field.
These $B_x$ and $B_z$ fields as well as the $B_y$ field
contribute to the dissipation of the shock.
Since these filaments and the magnetic field are never generated in 1D simulations,
the shock structure may differ significantly from those in 1D cases.

\begin{figure*}[htbp]
\plotone{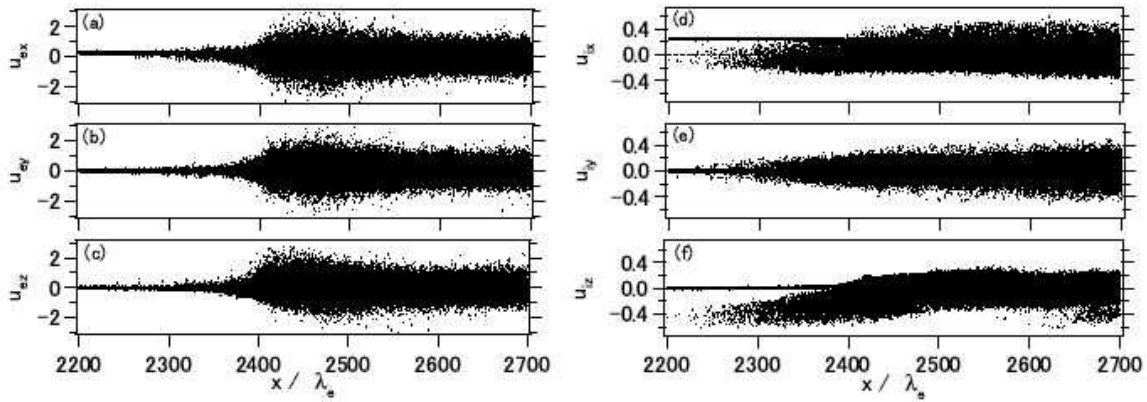}
\caption{
Phase-space plots of electrons (left panel) and ions (right panel).
The (from top to bottom) $x$, $y$, and $z$ components of the four velocities are shown
in each panel.
Both species are mostly thermalized within the shock transition region.
A fraction of the incoming ions are reflected at $x \sim 2400 \lambdae$.
}
\label{fig:phase_space} 
\end{figure*}

Figure \ref{fig:phase_space} shows phase-space plots of the electrons and the ions.
Here, each component of the four velocities
($u_j = \gamma v_j/c$,
where $j=x,y,z$ and $\gamma \equiv (1 - v^2/c^2)^{-1/2}$ is the Lorentz factor of the particle)
are plotted as a function of the $x$-coordinate.
Both electrons and ions from upstream
are mostly dissipated and isotropically thermalized through the transition region
($2300 \lambdae < x < 2450 \lambdae$).
It is observed that a fraction of ions are reflected at $x \sim 2400 \lambdae$
(i.e., the ramp)
and then gyrate back downstream with slight acceleration
forming the foot structure.
This is a well-known characteristic of supercritical shocks
and has been observed in many numerical simulations \citep[e.g.,][]{Leroy81, Burgess89}.
In contrast,
the electrons have
no prominent substructures in phase space.

\subsection{Foot dynamics}

\begin{figure}[htbp]
\plotone{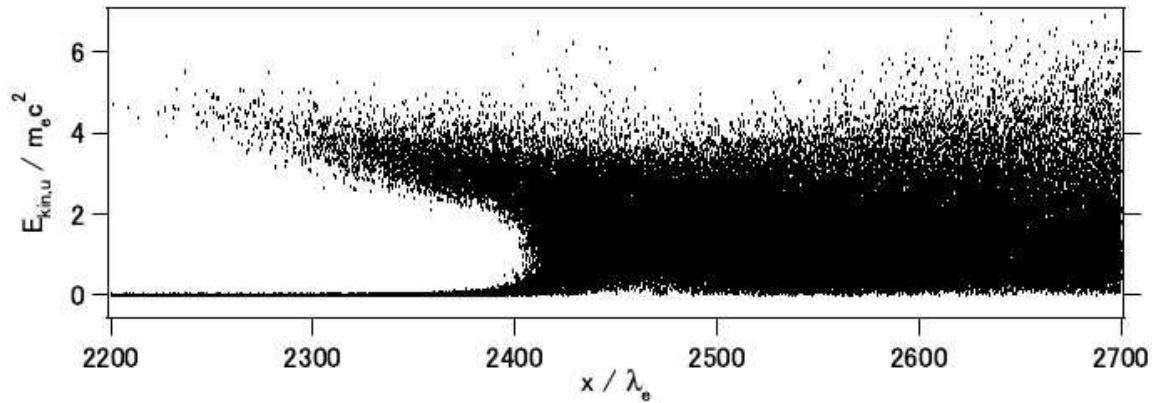}
\caption{
Kinetic energy of the ions measured in the upstream frame.
}
\label{fig:E_kin_u_ion}
\end{figure}

Figure~\ref{fig:E_kin_u_ion} shows
the distribution of 
the ion kinetic energy measured in the upstream frame,
$E_\mathrm{kin,u} = (\gamma_\mathrm{u} - 1)mc^2$, where $\gamma_\mathrm{u}$
is the particle Lorentz factor measured in the upstream frame,
as a function of the $x$-coordinate (in the downstream frame)
at $\omegape t = 8000$.
In this figure,
the incoming and reflected ions
in the foot region ($2200 < x/\lambdae < 2400$)
can be clearly distinguished from each other
using a threshold energy of, for example, $E_\mathrm{kin,u}/\me c^2 = 1$:
the reflected ions with $E_\mathrm{kin,u}/\me c^2 > 1$
and the incoming ions with $E_\mathrm{kim,u}/\me c^2 < 1$.
The reflected ions are further divided into two populations:
those streaming upstream measured in the shock rest frame
and those streaming downstream.
Thus,
we can investigate the foot dynamics on the basis of a simple fluid model
that consists of a single electron fluid and three ion fluids
(incoming ions, reflected ions streaming upstream,
and those streaming downstream),
which is similar to the model used in \citet{Leroy83}.
For convenience,
we denote the electrons, the incoming ions,
the reflected ions streaming upstream, and those
streaming downstream by the symbols e, I, R$_-$, and R$_+$,
respectively.

\begin{figure}[htbp]
\plotone{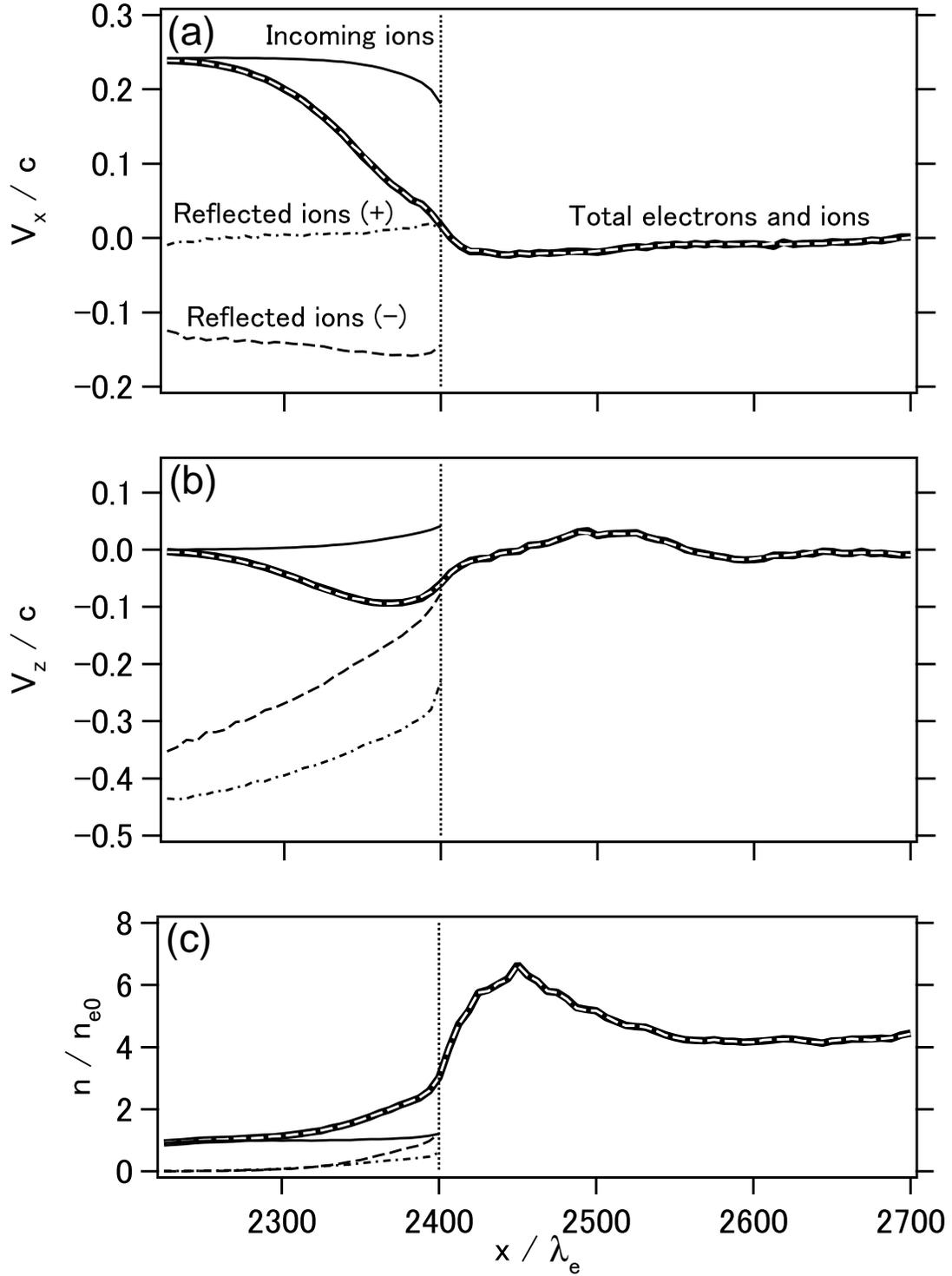}
\caption{
Profiles of the mean velocities in the (a) $x$-
and (b) $z$-directions,
and (c) the mean number densities normalized by
the upstream number density.
}
\label{fig:V_prof}
\end{figure}

Figures~\ref{fig:V_prof}(a) and (b) show the mean velocity of each fluid component
in the $x$- and $z$-directions
and Fig.~\ref{fig:V_prof}(c) shows the number densities
normalized by the upstream number density.
In the downstream region ($x > 2400 \lambdae$),
only the values for all the ions are shown
because classifying the ions
by the above method is meaningless in that region.
It shows that
the mean velocities and the number density of all electrons (thick curves)
and for all ions (dashed white curves)
agree well with each other in both the upstream and downstream regions,
indicating that the massless electron fluid model \citep{Leroy83}
holds well, at least on average, even in this high Mach number and
low ion-to-electron mass ratio case.

\subsubsection{Electrostatic instabilities and heating}
\begin{figure}[htbp]
\plotone{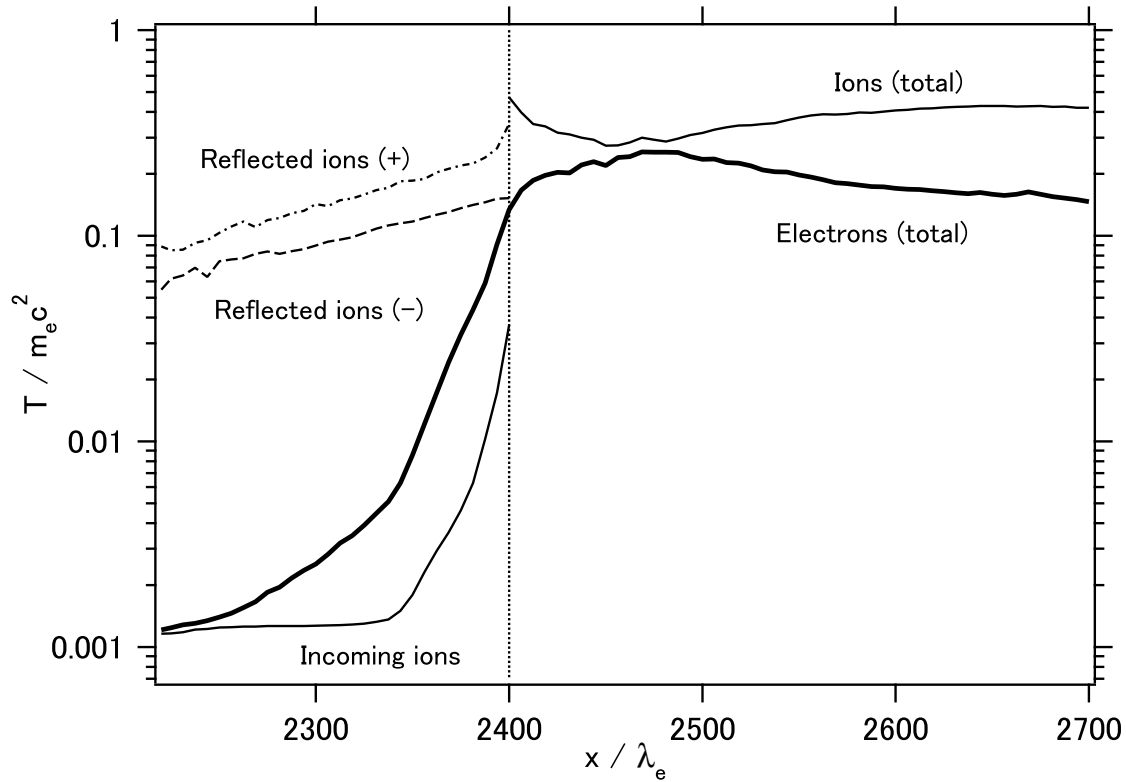}
\caption{
Temperature profile of each component.
Both incoming electrons (thick solid curve)
and incoming ions (thin solid curve) are
heated in the foot region ($x < 2400 \lambdae$).
}
\label{fig:T_prof}
\end{figure}

The local temperatures of the respective components were calculated using the mean velocities obtained above and they are plotted in
Fig.~\ref{fig:T_prof}.
It shows that in the foot region,
the electrons are heated in two steps:
the first step in $x \le 2350 \lambdae$ (region 1)
and the second step in $2350 \le x /\lambdae \le 2400$ (region 2).
The incoming ions are also heated in region 2.
This sequential electron heating process
together with the ion heating
suggests that
the model for very high Mach number shocks proposed by \citet{Papadopoulos88}
is valid in the foot region,
in which the incoming electrons are first heated by the Buneman instability \citep{Buneman58}
for reflected ions \citep{Auer71} and subsequently,
after the electrons have been heated to a certain temperature,
they are further heated
by the ion-acoustic instability for incoming ions.
The latter instability can also heat the ions.
This process has been studied by \citet{Cargill88} for $\MA \sim 50$ and $500$
with hybrid simulations with a phenomenological resistivity
and also by \citet{Shimada00} for $\MA \sim 10.5$
with 1D PIC simulations.

\begin{figure}[htbp]
\plotone{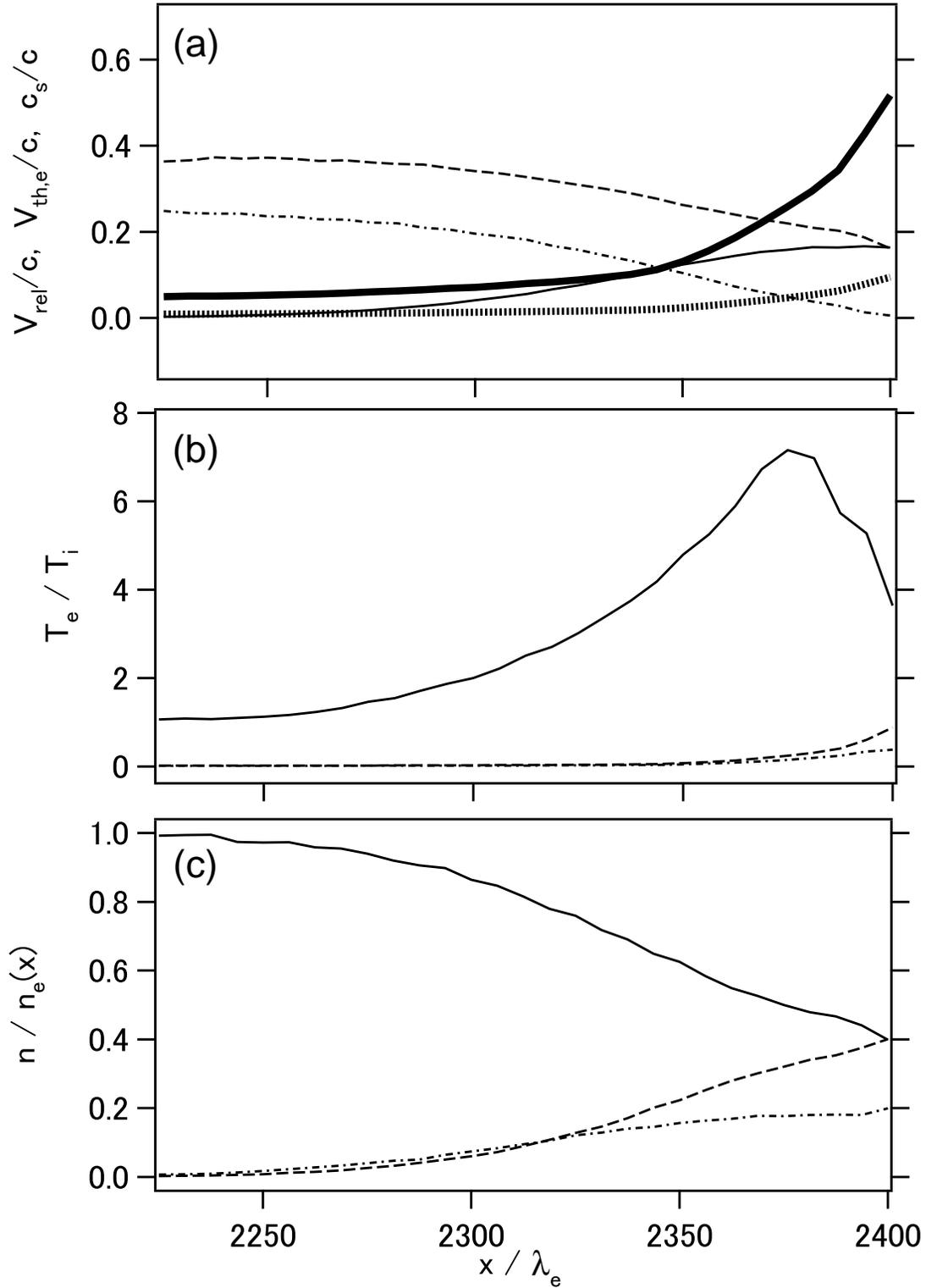}
\caption{
Profiles of the quantities in the foot region for
incoming ions (thin solid curves),
reflected ions streaming upstream (dashed curves),
and reflected ions streaming downstream (dot-dashed curves). 
(a) The velocities relative to the electron velocity in the $x$-direction (absolute values).
The electron thermal velocity (thick solid curve)
and the ion-acoustic speed (thick dotted curve) are also shown.
(b) The electron-to-ion temperature ratios.
(c) The number densities normalized by the local electron
number density, $n_\mathrm{e}(x)$.
}
\label{fig:Vrel_profiles}
\end{figure}

This heating process is expected to operate
in very high Mach number shocks and so
it should also operate in the present case ($\MA \sim 130$).
Figure~\ref{fig:Vrel_profiles} shows
several quantities of each ion component obtained from Figs.~\ref{fig:V_prof} and \ref{fig:T_prof}; specifically, it shows
profiles of the mean velocity relative to the electron velocity in the $x$-direction,
the electron-to-ion temperature ratio,
and the number density normalized by the local electron number density $n_\mathrm{e}(x)$.
In region 1 ($x < 2350 \lambdae$),
the reflected ions streaming upstream have
a significantly higher velocity relative to the electrons
than the electron thermal velocity.
On the other hand,
in region 2 ($x > 2350 \lambdae$),
the electron-to-ion temperature ratio for the incoming ions
increases to a large value
and also the incoming ion velocity relative to the electron velocity
becomes large
due to the large deceleration of the electrons [see Fig.~\ref{fig:V_prof}(a)]
so that it becomes higher than the ion-acoustic speed, $c_\mathrm{s} \equiv (\kB \Te/\mi)^{1/2}$.
These conditions are indeed preferable to
the Buneman instability in region 1
and the ion-acoustic instability in region 2.

Here,
we show
the instabilities that operate in the foot region by performing local linear analysis
with the fluid quantities (namely, the mean velocities, the number densities,
and the temperatures shown in Figs.~\ref{fig:V_prof} and \ref{fig:T_prof}).
Approximating the distribution of each component as a Maxwellian distribution,
\begin{equation}
	f_0^{(\mathrm{s})}(v_x, v_y, v_z)
	= \frac{\ns}{\pi^{3/2} \as^3}
		\exp \left[
		-\frac{(v_x - \Vsx)^2 + v_y^2 + (v_z-\Vsz)^2}{\as^2} \right],
\end{equation}
where $\mathrm{s} = (\mathrm{e, I, R_-, R_+})$,
$\ns$ is the number density, $\Vsx$ and $\Vsz$ are the streaming velocities
in the $x$- and $z$-directions respectively,
and $\as = (2 \kB \Ts/\ms)^{1/2}$ is the thermal velocity.
We solve the following dispersion relation for the electrostatic mode
with the wavevector in the $x$-direction:
\begin{equation}
	k_x^2 + 2\sum_\mathrm{s} \kDs^2 (1 + \zetas Z(\zetas)) = 0,
	\label{eq:es_disp_rel}
\end{equation}
where
\begin{equation}
	\kDs \equiv \omegaps / \as,
\quad
\omegaps \equiv \left( \frac{4\pi n_\mathrm{s} q_\mathrm{s}^2}{\ms} \right)^{1/2},
\end{equation}
and
\begin{equation}
	\zetas = \zetas(\omega, k_x) \equiv (\omega - k_x \Vs) / k_x \as.
\end{equation}
The function $Z(\zeta)$ is the plasma dispersion function \citep{Fried}
defined by
\begin{equation}
	Z(\zeta) \equiv \pi^{-1/2} \int_{-\infty}^{\infty} \frac{e^{-z^2}}{z - \zeta} dz.
\end{equation}
Table~\ref{table:sim_values} summarizes some quantities used in the following analysis.
\begin{table*}[!h]
\caption{
Quantities obtained from the simulation.
The units for the $x$-coordinate, the velocity, the number density, and the temperature
are $\lambdae$, $c$, $\nez$, and $\me c^2$, respectively.
}
\begin{center}
\scalebox{0.85}{  
\begin{tabular}{ccccccccccccccccc}
	\hline
	$x$ &
	$V_{\mathrm{e},x}$ & $V_{\mathrm{e},z}$ & $n_\mathrm{e}$ & $T_\mathrm{e}$ &
	$V_{\mathrm{I},x}$ & $V_{\mathrm{I},z}$ & $n_\mathrm{I}$ & $T_\mathrm{I}$ &
	$V_{\mathrm{R-},x}$ & $V_{\mathrm{R-},z}$ & $n_\mathrm{R-}$ & $T_\mathrm{R-}$ &
	$V_{\mathrm{R+},x}$ & $V_{\mathrm{R+},z}$ & $n_\mathrm{R+}$ & $T_\mathrm{R+}$\\
	\hline
	$2300$ &
	$0.20$ & $-0.042$ & $1.1$ & $2.5\times 10^{-3}$ &
	$0.24$ & $3.3\times 10^{-3}$ &$0.99$ & $1.3\times 10^{-3}$ &
	$-0.14$ & $-0.27$ & $0.069$ & $0.090$ &
	$4.6 \times10^{-3}$ & $-0.39$ & $0.084$ & $0.14$\\
	$2350$ &
	$0.11$ & $-0.089$ & $1.6$ & $8.6\times 10^{-3}$ &
	$0.23$ & $0.013$ & $1.0$ & $1.8\times 10^{-3}$ &
	$-0.15$ & $-0.19$ & $0.37$ & $0.12$ &
	$5.5 \times10^{-3}$ & $-0.34$ & $0.26$ & $0.19$\\
	$2375$ &
	$0.062$ & $-0.093$ & $2.1$ & $0.033$ &
	$0.22$ & $0.023$ & $1.0$ & $4.6\times 10^{-3}$ &
	$-0.16$ & $-0.15$ & $0.68$ & $0.14$ &
	$0.013$ & $-0.32$ & $0.38$ & $0.22$\\
	\hline
\end{tabular}
} 
\end{center}
\label{table:sim_values}
\end{table*}
We use here the dispersion relation for unmagnetized plasmas
given by Eq.~(\ref{eq:es_disp_rel})
instead of that for magnetized plasmas
because the magnetic field is sufficiently weak in the present case;
the condition for the unmagnetized approximation is given by
$k^2 \gg k_\mathrm{cs}^2$ where $k_\mathrm{cs} \equiv \sqrt{2} |\omegacs| / \as$;
in other words, for all species,
the wavelength is much smaller than
the Larmor radius defined for the thermal velocity.
As shown below, the wavenumbers of the instabilities are typically $kc/\omegape > 3$
and $k_\mathrm{cs}c/\omegape$ is $\sim 0.5$ for electrons, $\sim 0.15$ for
incoming ions, and $\sim 0.01$ for reflected ions in the foot region.
Therefore,
the unmagnetized approximation can be used in this case.

\begin{figure}[htbp]
\plotone{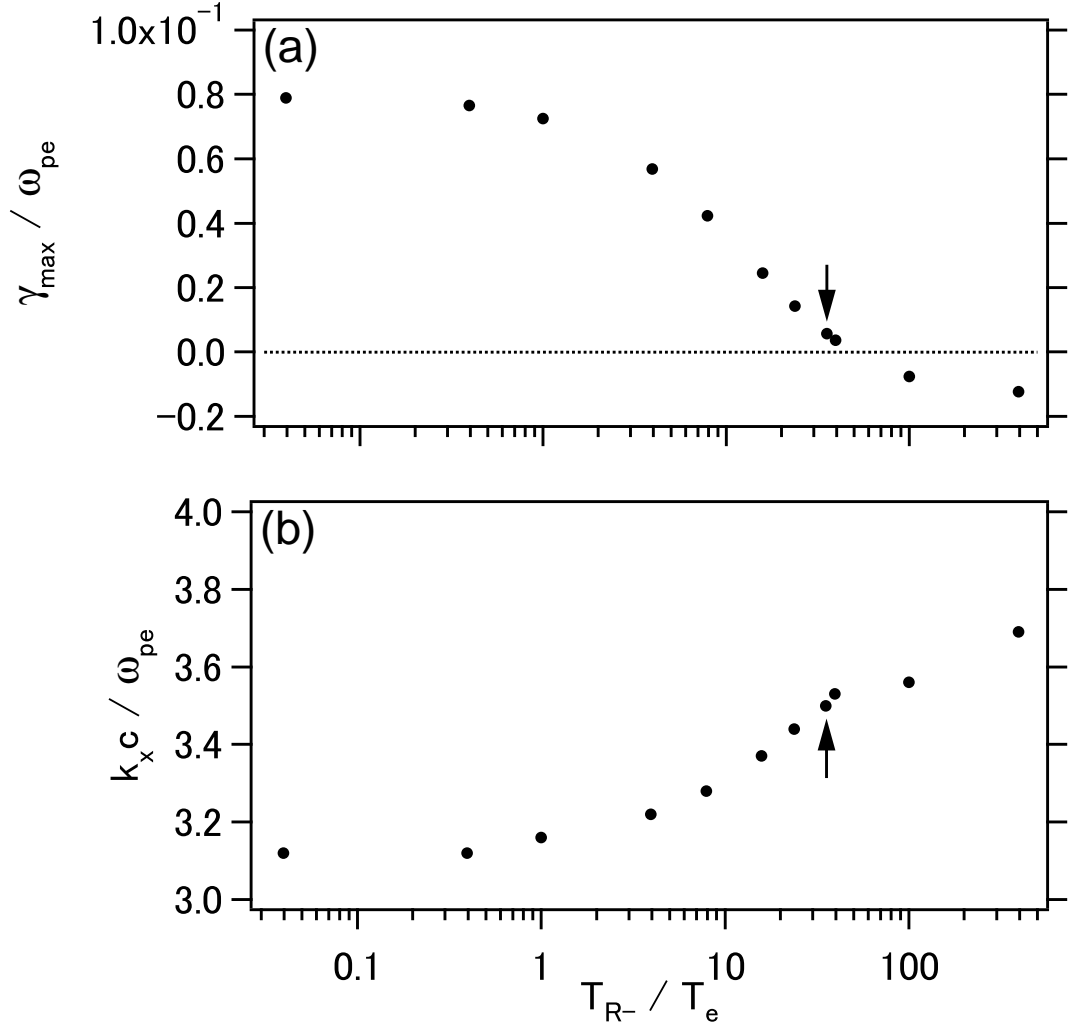}
\caption{
(a) Maximum linear growth rates of the electrostatic mode
and
(b) the wave numbers at the maximum growth rates
as functions of the temperature of the reflected ions
streaming upstream, $T_{R-}$,
calculated for the quantities obtained from the simulation at $x = 2300 \lambdae$
(except $T_{R-}$).
Those for the original value of $T_{R-}$
in the simulation are indicated by the arrows.
}
\label{fig:BI_x2300}
\end{figure}

When performing the linear analysis with the local quantities at $x = 2300 \lambdae$,
we found an unstable electrostatic mode
whose wavenumber ($k_x \sim 3.5 \omegape/c$) and
frequency ($\omega' \sim 0.048 \omegape$
in the rest frame of the reflected ions $R_-$)
are similar to those of the Buneman instability
between electrons and reflected ions streaming upstream
($k_x \sim 2.9 \omegape/c$ and $\omega' \sim 0.05 \omegape$ in the rest frame of $R_-$).
However, the obtained growth rate ($\gamma \sim 0.0057 \omegape$) 
is one order of magnitude smaller than
the typical growth rate of the Buneman instability ($\gamma \sim 0.087 \omegape$).
This is because of the relatively high temperature of the reflected ions streaming upstream, $T_{R-}$,
as is shown in Fig.~\ref{fig:T_prof},
while the ordinary Buneman instability assumes that both species are cold.
Figure~\ref{fig:BI_x2300} shows
the maximum linear growth rates of this mode
together with their wavenumbers
calculated for the quantities at $x = 2300 \lambdae$
while varying $T_{R-}$.
When $T_{R_-}$ approaches zero,
the growth rate becomes large
and approaches to a typical value for the Buneman instability.
Thus,
we regard this mode as a Buneman instability
between the electrons and the reflected ions streaming upstream
with a reduction in the growth rate
due to the relatively high temperature of the reflected ions.

On the other hand,
we found another unstable electrostatic mode at $x \sim 2350 \lambdae$.
This mode has a maximum growth rate $\gamma_\mathrm{max} \sim 0.02 \omegape$
and a frequency $\omega \sim 0.52 \omegape$ at $k_x \sim 4.85 \omegape/c$.
This leads to a phase velocity of $\sim -0.017c$ in the incoming ion
rest frame.
For the same parameters,
the dispersion relation of the ion-acoustic instability \citep{Ichimaru}
between electrons and incoming ions
gives $\gamma_\mathrm{max} \sim 0.015 \omegape$ at $k_x \sim 4.0 \omegape/c$
and a phase speed of $c_\mathrm{s} \sim -0.016c$
in the incoming ions rest frame.
Both agree well with each other and
thus we regard this mode as an ion-acoustic instability
between electrons and incoming ions.

\begin{figure}[htbp]
\plotone{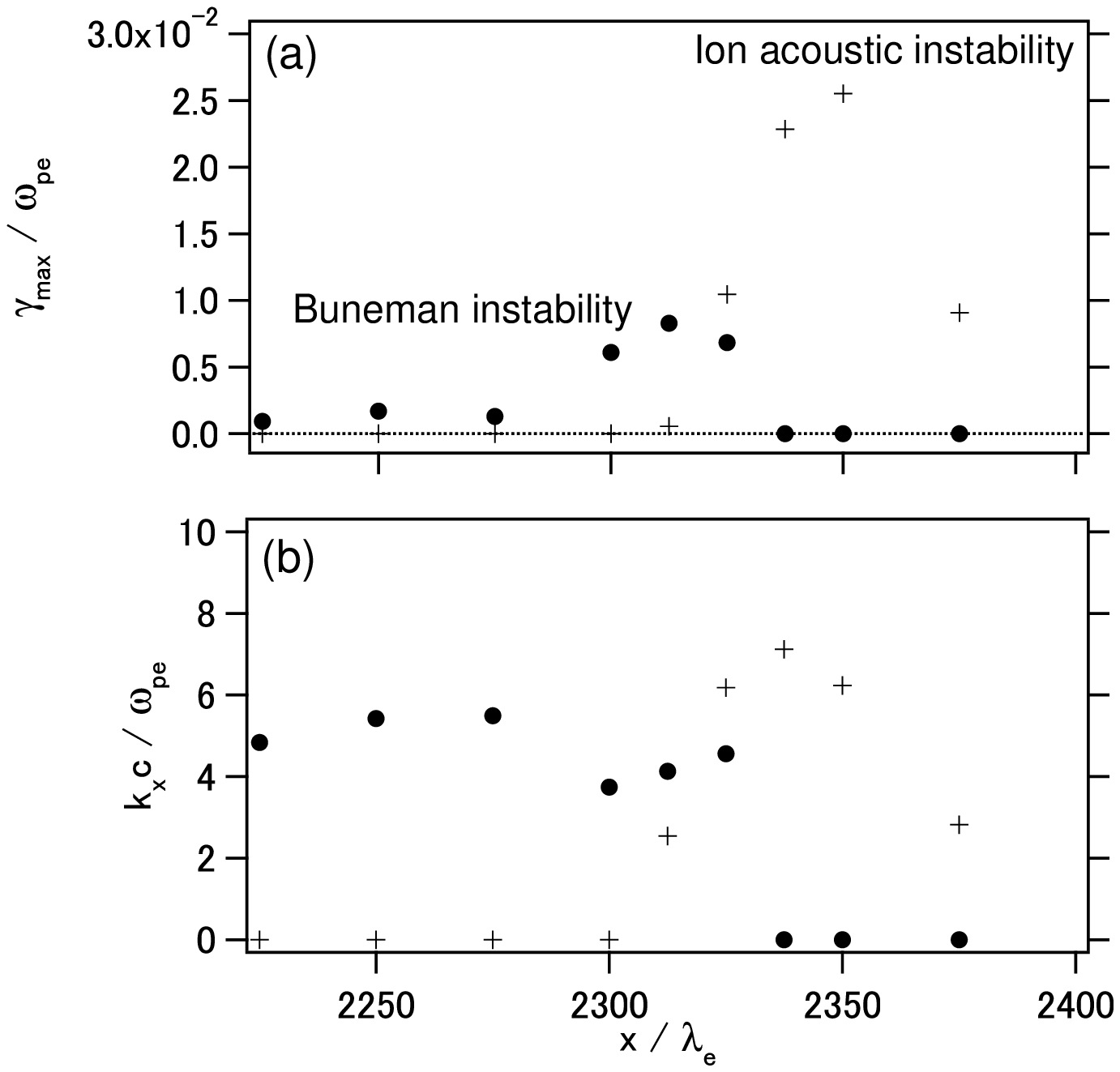}
\caption{
(a) Maximum linear growth rates of the Buneman instability (filled circles)
and the ion-acoustic instability (crosses) as functions of $x$.
(b) The wave numbers at the maximum growth rates
for the respective instabilities.
}
\label{fig:BI_IA_gr}
\end{figure}

Figure~\ref{fig:BI_IA_gr} summarizes the results for this local linear analysis
over the foot region.
The Buneman instability develops upstream of the foot region ($x \le 2325 \lambdae$), whereas the ion-acoustic instability dominates downstream of the foot region ($x \ge 2325 \lambdae$).
This feature is consistent with the evolution of
the electron and incoming ion temperatures shown in Fig.~\ref{fig:T_prof};
the electrons are first heated by the Buneman instability
and then both electrons and incoming ions are heated by the ion-acoustic instability.
Note that there is a region where both instabilities can coexist
($2312 \le x / \lambdae \le 2325$).

\begin{figure}[htbp]
\plotone{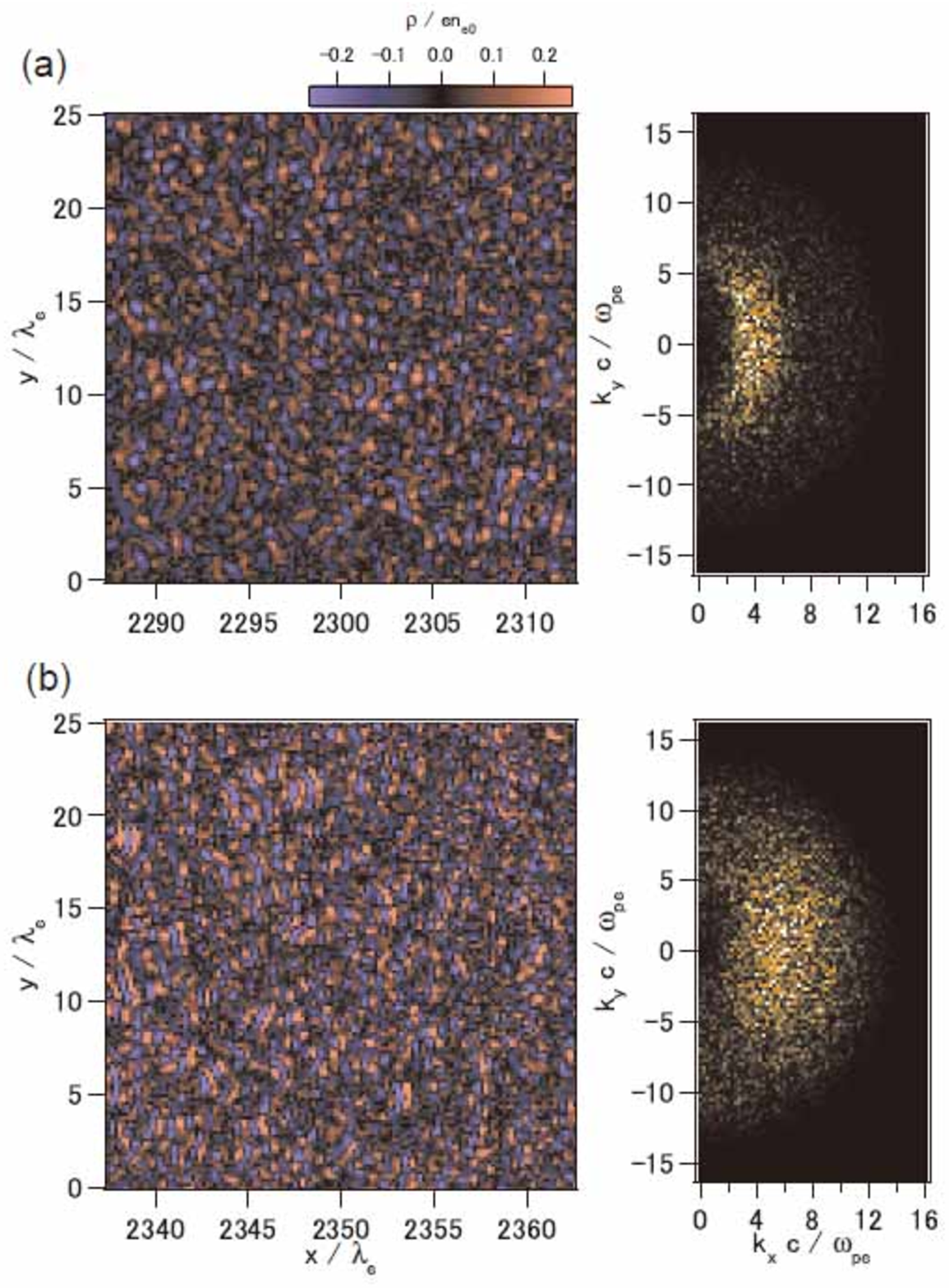}
\caption{
Charge density (left panels) and its power spectrum (right panels):
around (a) $x \sim 2300 \lambdae$
and (b) $x \sim 2350 \lambdae$.
The portions where the power is strong are consistent with
the linear theory of (a) the Buneman instability 
and (b) the ion-acoustic instability, respectively.
}
\label{fig:rho_power}
\end{figure}

Since both the Buneman and the ion-acoustic instabilities are electrostatic modes,
they are always associated with the charge density $\rho$
and can be investigated through it.
Figures \ref{fig:rho_power}(a) and (b) show
the charge density and its power spectrum in two rectangular areas in the foot region, namely
$x \sim 2300 \lambdae$ (where the Buneman instability dominates)
and $x \sim 2350 \lambdae$ (where the ion-acoustic instability dominates), respectively.
The peak positions of these power spectra agree well 
with the wavenumbers for the maximum growth rates obtained by
the linear theory shown in Fig.~\ref{fig:BI_IA_gr}.
Note that both spectra are not concentrated on the $k_x$-axis
but extend in the $k_y$-direction.
This results in the wavy appearance of both modes
in real space (left panels) and
is a well-known characteristic of both instabilities
in multiple dimensions.

\subsubsection{Filamentary structures}
\begin{figure}[htbp]
\plotone{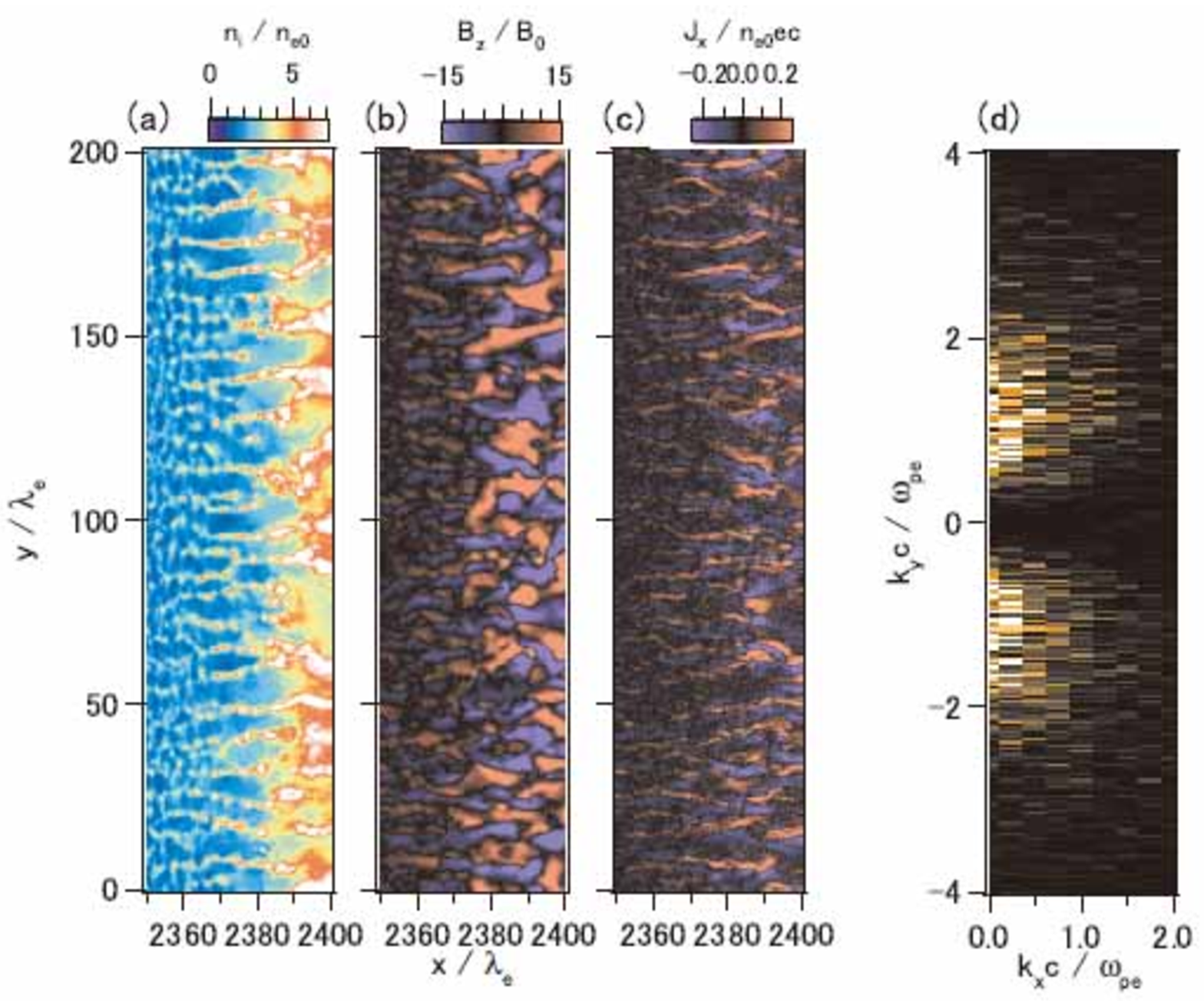}
\caption{
(a) Ion number density, (b) magnetic field $B_z$, and
(c) current density in the $x$-direction, $J_x$, around $x \sim 2375 \lambdae$.
(d) Power spectrum of $J_x$ calculated in the region $2362.5 < x/\lambdae < 2385.5$.
}
\label{fig:foot_filaments}
\end{figure}

As mentioned above,
the ion number density
in the foot region (Fig.~\ref{fig:ni_160000}) contains many filamentary structures.
Figure~\ref{fig:foot_filaments} shows that
these filaments are associated with current filaments and filamentary magnetic fields.
These filaments are similar to those observed in unmagnetized shocks,
which are generated by the beam-Weibel instability.
In our previous papers, we showed that
the ion beam--Weibel instability develops and generates currents filaments
even for nonrelativistic flow speeds \citep{Kato08, Kato10}.
Therefore,
it is plausible that these filaments are generated by the ion beam-Weibel instability.
To confirm this,
we performed linear analysis
in the same manner as that used to obtain Fig.~\ref{fig:BI_IA_gr}
except that
we here consider the electromagnetic modes with wavevectors in the $y$-direction.
In the present case,
since the wavenumber is too low to employ the unmagnetized approximation
for electrons,
we solve the following dispersion relation
in the electron rest frame, which
includes the effect of the magnetic field for the electrons
(the ions are assumed to be unmagnetized):
\begin{equation}
	\mathrm{det}\Lambda = 0,
\end{equation}
where
\begin{eqnarray}
\Lambda_{xx}
	&=& 1 - \left(\frac{kc}{\omega}\right)^2
	+ \frac{1}{2} \left(\frac{\omegape}{\omega}\right)^2 \xi_0 \left[Z(\xi_1) + Z(\xi_{-1})\right]
	+ \sum_\mathrm{s} \left( \frac{\omegaps}{\omega} \right)^2 \left[ \alps + 2\tVsx^2(1+\alps)\right],\\
\Lambda_{yy}
	&=& 1 + 2\left(\frac{\omegape}{k a_\mathrm{e}}\right)^2 \left[1 + \xi_0 Z(\xi_0)\right]
	+ 2 \sum_\mathrm{s} \left( \frac{\omegaps}{k\as} \right)^2 (1 + \alps),\\
\Lambda_{zz}
	&=& 1 - \left(\frac{kc}{\omega}\right)^2
	+ \frac{1}{2} \left(\frac{\omegape}{\omega}\right)^2 \xi_0 \left[Z(\xi_1) + Z(\xi_{-1})\right]
	+ \sum_\mathrm{s} \left( \frac{\omegaps}{\omega} \right)^2 \left[ \alps + 2\tVsz^2(1+\alps)\right],\\
\Lambda_{xy}
	&=& \Lambda_{yx} = 2 \sum_\mathrm{s} \left( \frac{\omegaps}{\omega} \right)^2 \tVsx \etas (1 + \alps),\\
\Lambda_{yz}
	&=& \Lambda_{zy} = 2 \sum_\mathrm{s} \left( \frac{\omegaps}{\omega} \right)^2 \tVsz \etas (1 + \alps),\\
\Lambda_{xz}
	&=& -\frac{i}{2} \left(\frac{\omegape}{\omega}\right)^2 \xi_0 \left[ Z(\xi_1) - Z(\xi_{-1}) \right]
	+ 2 \sum_\mathrm{s} \left( \frac{\omegaps}{\omega} \right)^2 \tVsx \tVsz (1 + \alps),\\
\Lambda_{zx}
	&=& \frac{i}{2} \left(\frac{\omegape}{\omega}\right)^2 \xi_0 \left[ Z(\xi_1) - Z(\xi_{-1}) \right]
	+ 2 \sum_\mathrm{s} \left( \frac{\omegaps}{\omega} \right)^2 \tVsx \tVsz (1 + \alps),
\end{eqnarray}
with
\begin{eqnarray}
	&&\xi_n \equiv \frac{\omega - n\omegace}{k a_\mathrm{e}},
	\quad
	\etas \equiv \frac{\omega}{k\as},
	\quad
	\alps \equiv \etas Z(\etas),\\
	&&\tVsx \equiv V_{\mathrm{s},x} / \as,
	\quad
	\tVsz \equiv V_{\mathrm{s},z} / \as.
\end{eqnarray}
In the above dispersion relation,
the sums run only for the ion species,
that is for $\mathrm{s} = \mathrm{I, R_-, R_+}$.

\begin{figure}[htbp]
\plotone{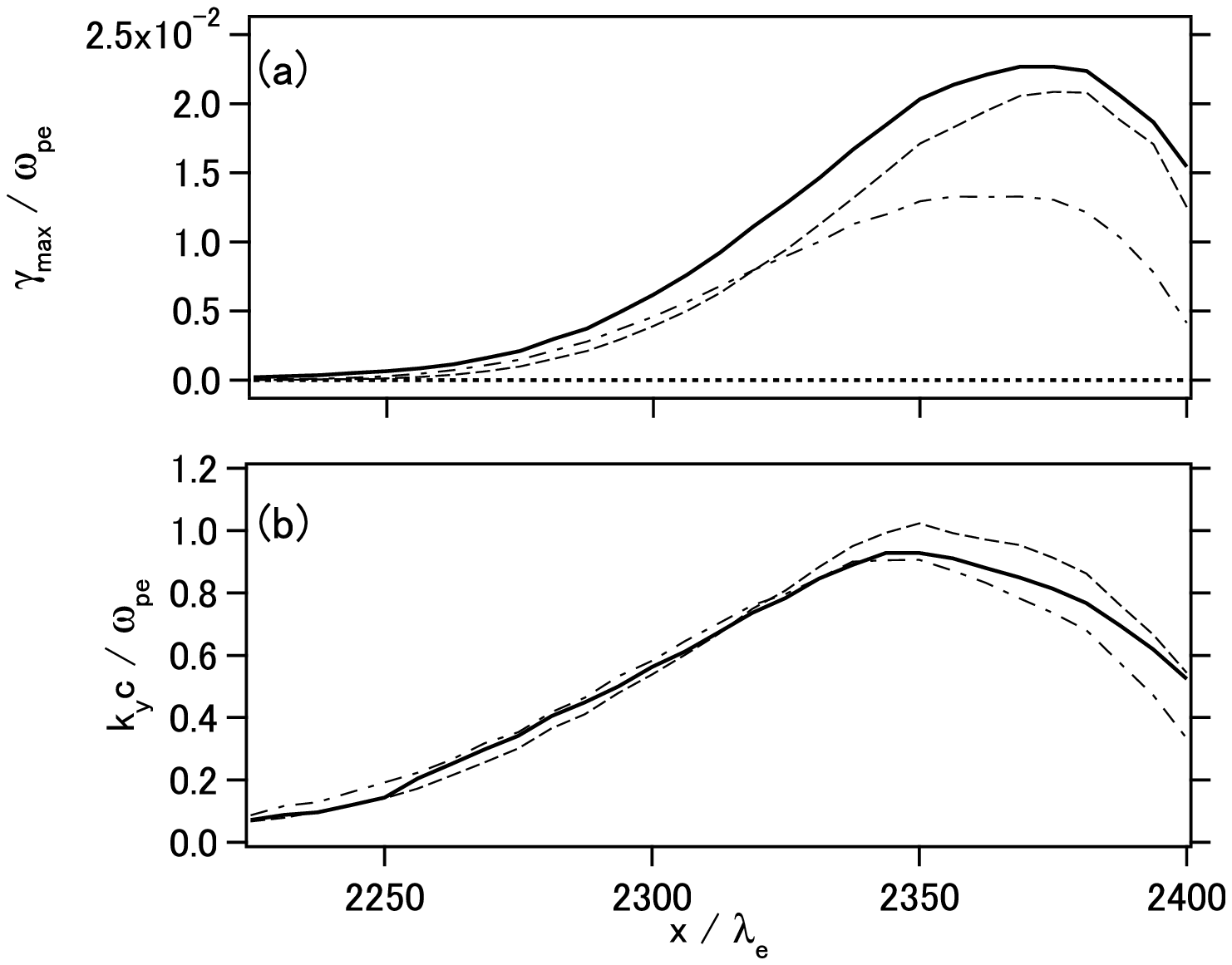}
\caption{
(a) The linear growth rate
and (b) wavenumber of the most unstable mode
of the electromagnetic instability
with the wave vectors parallel to the background magnetic field
(i.e., in the $y$-direction) (solid curves).
The real frequency is shown by the dotted curve in the panel (a), which
shows that the mode is a purely growing mode.
Those for the ion beam-Weibel instability are shown by
the dashed curves (for $\tVsx$) and dot-dashed curves (for $\tVsz$),
respectively.
}
\label{fig:beam_Weibel_gr}
\end{figure}

The results are shown in Fig.~\ref{fig:beam_Weibel_gr} by the solid curves.
The mode is unstable in the foot region
and it grows at a comparable growth rate to
those of electrostatic modes (see Fig.~\ref{fig:BI_IA_gr}).
The wavenumber obtained in the linear analysis
near $x=2360\lambdae$ is typically $k_y \sim 0.8$;
this value agrees well with the simulation result shown in Fig.\ref{fig:foot_filaments}(d).
Note that the real frequency of the mode (dotted curve) is zero; that is,
it is a purely growing mode.

This mode can be regarded as an ion beam-Weibel instability.
Indeed,
as shown in Fig.~\ref{fig:beam_Weibel_gr},
the maximum growth rate and the wave number essentially
agree with those obtained from the dispersion relation for the beam-Weibel
instability using the unmagnetized approximation:
\begin{equation}
	\omega^2 - (kc)^2 + \sum_\mathrm{s} \omegaps^2
	\left[ \alps + 2 \tVs^2 (1 + \alps) \right]
	= 0,
\end{equation}
where $\tVs$ is taken to be either $\tVsx$ (shown by the dashed curves)
or $\tVsz$ (the dot-dashed curves)
and $\mathrm{s} = \mathrm{e, I, R_-, R_+}$.
Thus,
it can be concluded that
the filamentary structure in the foot region
is generated by the ion beam-Weibel instability. 
The strong magnetic field generated by the instability
would contribute to the thermalization of the incoming ions
immediately upstream of the ramp (see Fig.~\ref{fig:T_prof}).

\subsection{Downstream temperature and jump condition}
In the downstream region,
we obtain a temperature ratio of $\Te/\Ti \sim 0.38$
from Fig.~\ref{fig:T_prof}.
Thus, the ratio is significantly smaller than unity,
although it is still much larger than those observed in several SNRs
[e.g., $\Te/T_\mathrm{p} < 0.07$ in SN1006; \citet{Ghavamian02}.]

\begin{figure}[htbp]
\plotone{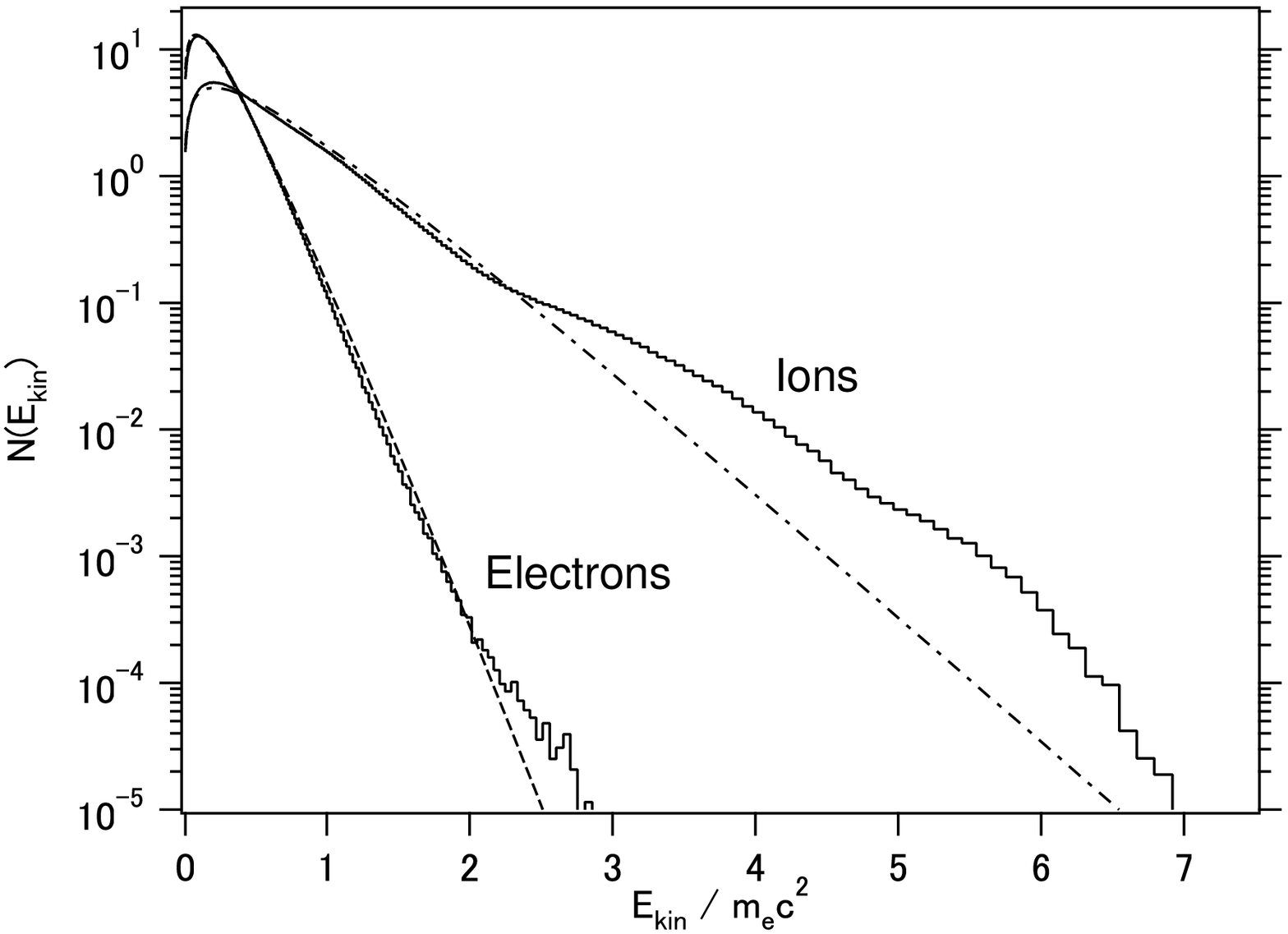}
\caption{
Kinetic energy distributions of the electrons and the ions
measured in the downstream frame (solid histograms)
normalized by the electron rest mass energy
in the downstream region within $2624 \lambdae < x < 2656 \lambdae$.
The dashed and dotted curves are the Maxwellian distributions with temperatures of
$\kB \Te / \me c^2 = 0.14$ for electrons
and $\kB \Ti / \me c^2 = 0.42$ for ions, respectively.
}
\label{fig:hist_d}
\end{figure}

Figure \ref{fig:hist_d} shows the kinetic energy distributions of
the electrons and the ions in a rectangular downstream region ($2624 \lambdae < x < 2656 \lambdae$).
Both distributions are fitted very well with the (3D and relativistic) Maxwellian distributions
\citep[e.g.,][]{Landau}
\begin{equation}
	f(\gamma) d\gamma \propto \gamma (\gamma^2-1)^{1/2} \exp(-\gamma mc^2 / \kB T ) d\gamma
\end{equation}
with temperatures $\kB \Te / \me c^2 = 0.14$ (for the electrons)
and $\kB \Ti / \me c^2 = 0.42$ (for the ions)
for $\Ekin / \me c^2 < 2$.
These temperatures again give a low temperature ratio of $\Te/\Ti \sim 0.33$. 
When the upstream bulk kinetic energy is completely dissipated
into thermal energy and the electron and the ions are in equipartition,
the temperature is given by
$\kB T / \me c^2 \sim 0.32$ for both species.
On the other hand,
when the electrons and the ions are thermalized separately,
their temperatures are $\kB \Te / \me c^2 = 0.021$
and $\kB \Ti / \me c^2 = 0.63$, respectively.
The ion distribution has a suprathermal tail
for $\Ekin / \me c^2 > 2$.
As shown in the next subsection,
these suprathermal ions originate from the reflected ions in the foot region.
In contrast, neither a suprathermal tail nor an accelerated population is clearly observed in the electron distribution.

Using the above results,
the shock jump conditions are calculated as follows.
In the shock rest frame,
the upstream flow velocity $V_1$
and the downstream flow velocity $V_2$ are given by
$V_1 = 0.33c$ and $V_2 = 0.084c$, respectively.
Thus, we have $V_1 / V_2 \sim 3.9$,
$N_2 / N_1 \sim 4.1$,
and $(\kB (\Te + \Ti) / \mi)^{1/2} \sim 0.14 c$,
where $N_1 (=\nez)$ and $N_2$ are the number densities in the upstream 
and downstream regions, respectively.
On the other hand,
the MHD Rankine--Hugoniot relations \citep[e.g.,][]{Tidman}
give $V_1 / V_2 = N_2 / N_1 \sim 4$
and $(\kB (\Te + \Ti) / \mi)^{1/2} \sim 0.15 c$
in the high Mach number limit.
Hence, the simulation results agree very well with the MHD Rankin--Hugoniot relations.
Even although,
in the downstream region,
the magnetic field reaches $\sim 15$ times the upstream fields
[see Fig.~\ref{fig:profiles}(b)],
the plasma beta is still high ($\beta \sim 25$) and 
the magnetic pressure is negligible for the jump condition.

\subsection{Acceleration of reflected ions}
\begin{figure}[htbp]
\includegraphics[width=14cm]{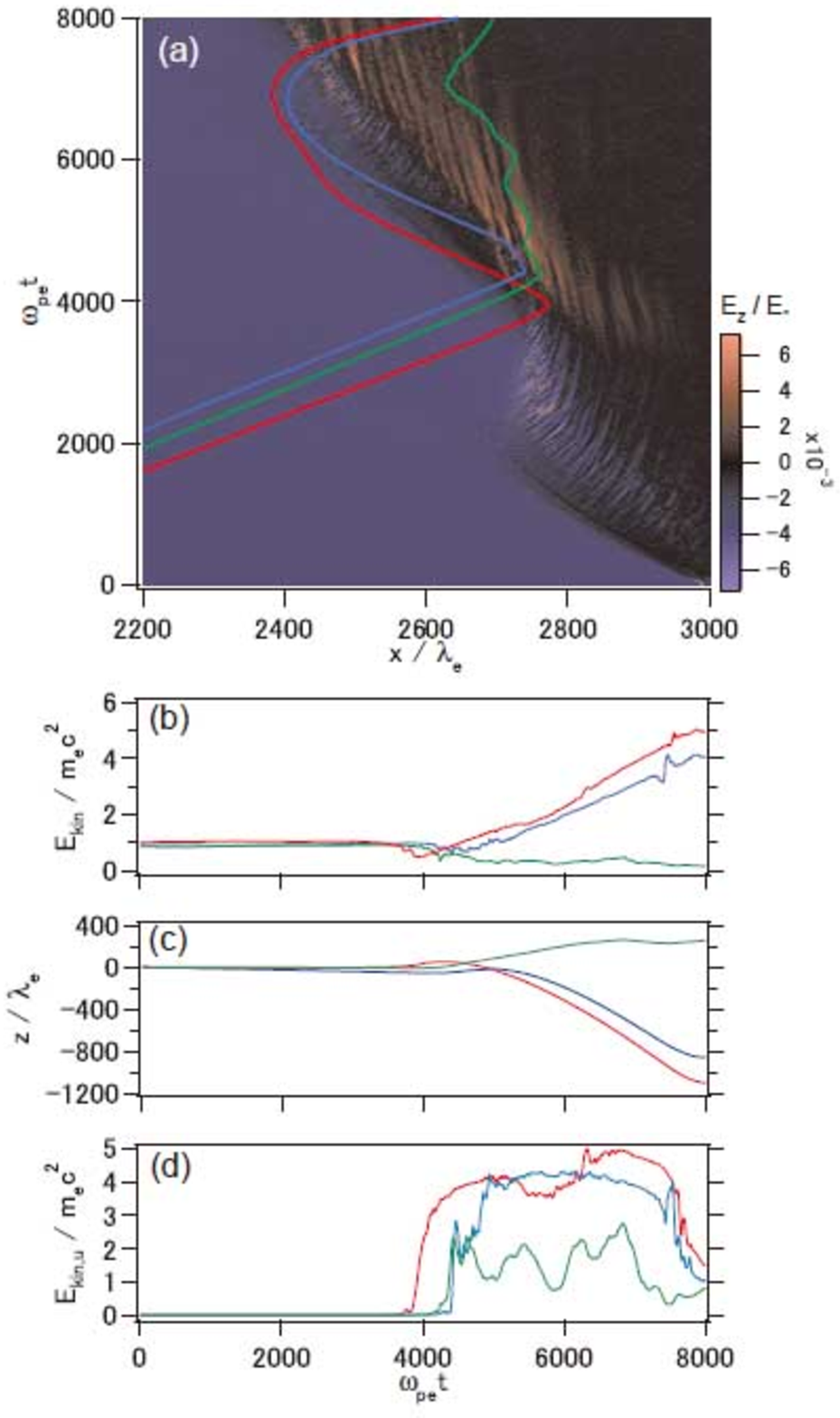}
\caption{
Histories of two accelerated ions (red and blue curves)
and one non-accelerated ion (green curve):
(a) Trajectories on the evolution of $E_z$.
(b) Kinetic energies of the ions measured in the downstream frame.
(c) $z$-coordinates of the ions.
(d) Kinetic energies of the ions measured in the upstream frame.
}
\label{fig:HE_history}
\end{figure}

As Fig.~\ref{fig:hist_d} shows,
a fraction of the ions are slightly accelerated
to $\Ekin/\me c^2 \sim 3 - 6$ (measured in the downstream frame)
and form suprathermal populations.
Figure~\ref{fig:HE_history}(a) shows the trajectories of
two typical accelerated ions (red and blue curves)
together with that of a non-accelerated ion (green curve).
It is clear that accelerated ions are reflected at the shock front
(i.e., the ramp) and go around the upstream region,
whereas non-accelerated ions are directly transmitted downstream.
Figure~\ref{fig:HE_history}(b) shows the kinetic energy history
of the three ions measured in the downstream frame.
It reveals that the kinetic energies of the ions
increase while they are in the upstream region.
This is simply acceleration by the motional electric field
$E_z = -V_x B_y/c$
in the downstream frame 
while they are gyrating in the foot region
\citep{Auer71,Phillips72}
after specular reflection at the shock front 
\citep{Paschmann82, Gosling82, Schwartz83}.
As Fig.~\ref{fig:HE_history}(c) shows,
the ions are accelerated when propagating in the $-z$-direction.
This process can be understand more clearly
in the upstream frame where there is (essentially) only a background
magnetic field and no motional electric field.
Figure~\ref{fig:HE_history}(d) shows the same kinetic energy histories
as Fig.~\ref{fig:HE_history}(b)
but measured in the upstream frame.
It shows that the ions gain energy at the reflection
and subsequently their kinetic energy remains almost constant.
Thus,
the ion acceleration is simply due to reflection at the ramp.

\subsection{Currents and magnetic field}

\begin{figure*}[htbp]
\plotone{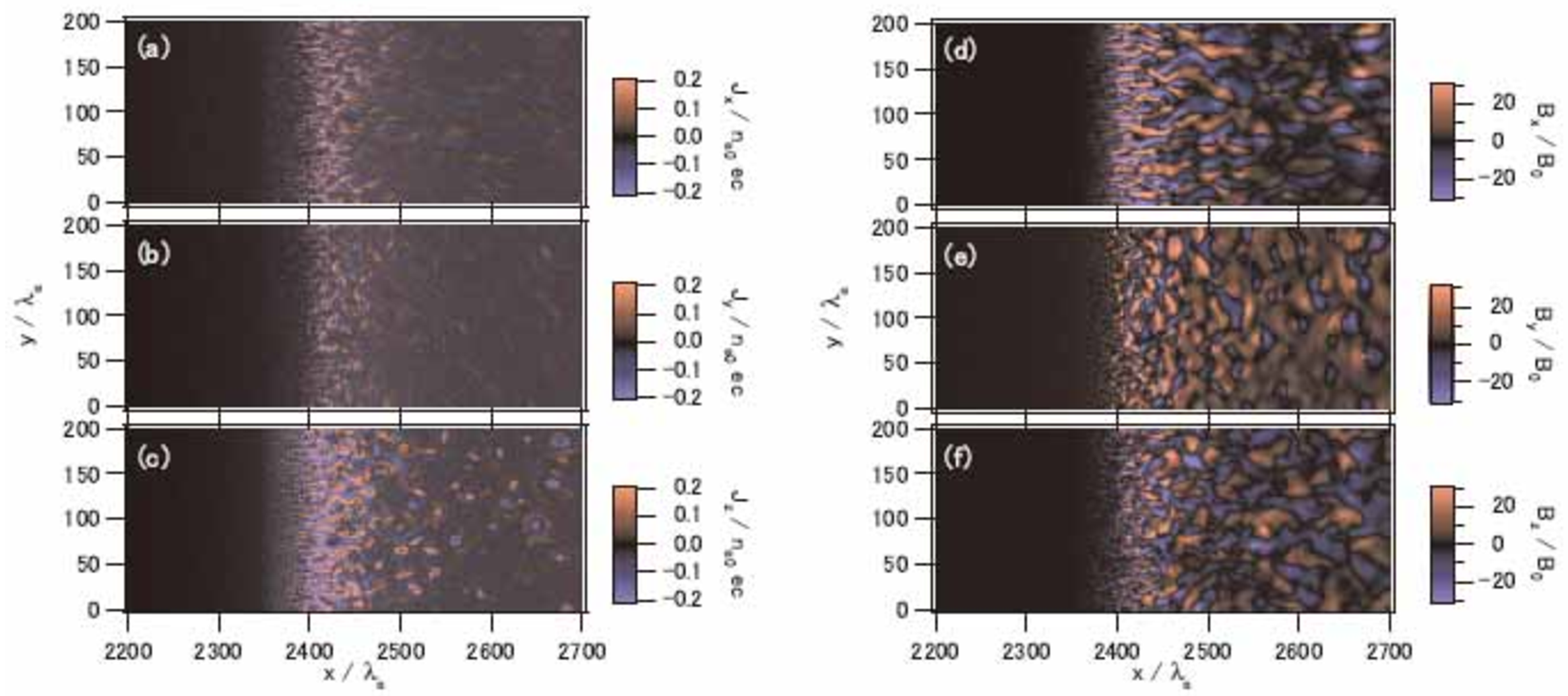}
\caption{
Left panel: each component of the current density
(a) $J_x$, (b) $J_y$, and (c) $J_z$
normalized by $\nez e c$.
Right panel: each component of the magnetic field
(d) $B_x$, (e) $B_y$, and (f) $B_z$
normalized by $B_0$.
There are numerous current filaments in the shock transition region and the downstream region
and they generate the magnetic field.
}
\label{fig:J_B}
\end{figure*}

Figure \ref{fig:J_B} shows each component of the current density
and the magnetic field.
The upstream background field, which is in the $y$ direction, is compressed in the shock transition region
as in the 1D simulations, although it fluctuates considerably in the present case.
As discussed above, many current filaments exist in the foot region of the
shock in $J_x$ and $J_z$;
the filamentary structures observed in the ion number density in Fig. \ref{fig:ni_160000}
indicate the presence of these current filaments.
The filaments generate a magnetic field
in the same way as a Weibel-mediated shock in unmagnetized plasmas
except that in the present cases
$B_x$ and $B_y$ components are generated by the current filaments in addition to the $B_z$ component
because the background field deflects the particles
in the $z$-direction and the current filaments can have a $J_z$ component as well as a $J_x$ component.

\begin{figure}[htbp]
\plotone{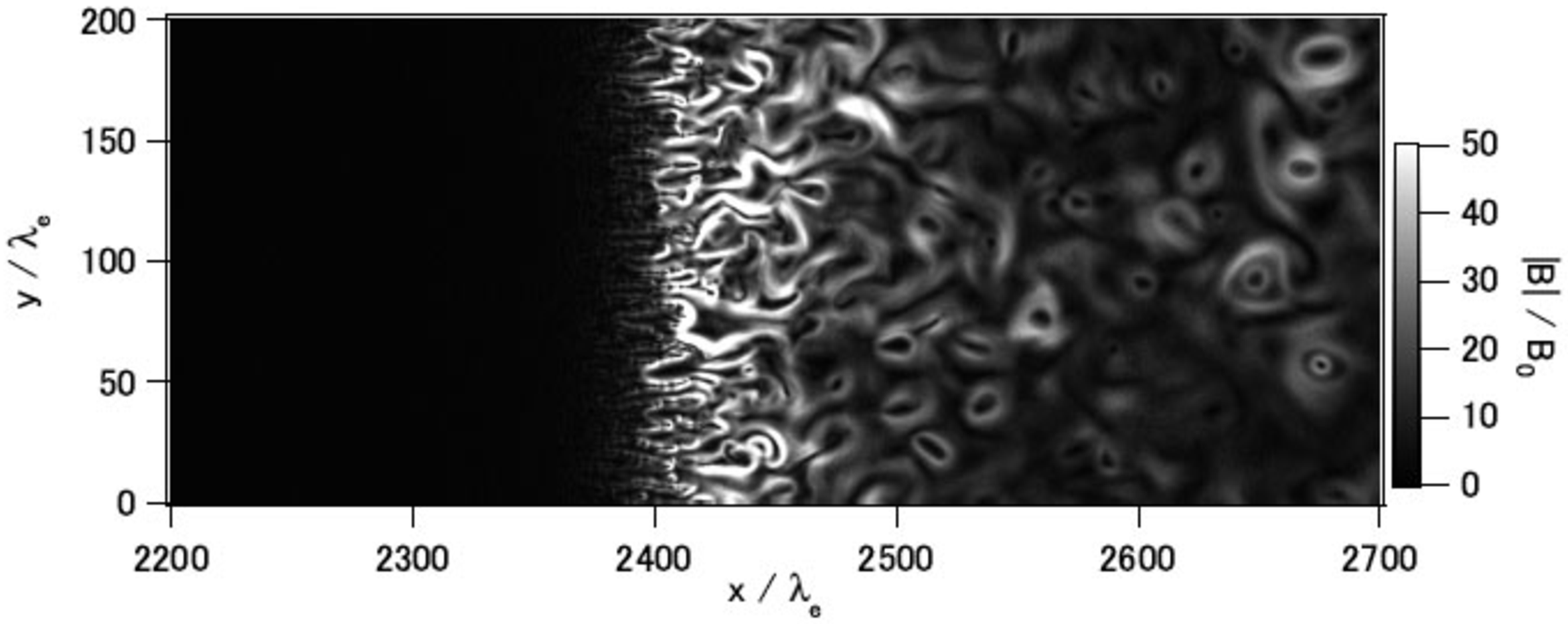}
\caption{
Magnetic field strength normalized by the upstream background magnetic field,
$|B| / B_0$. There is a strong, highly tangled magnetic field in both the shock transition region
($|B|/B_0 \sim 40$) and the downstream region ($|B|/B_0 \sim 15$).
}
\label{fig:B_abs_map}
\end{figure}

\begin{figure*}[htbp]
\plotone{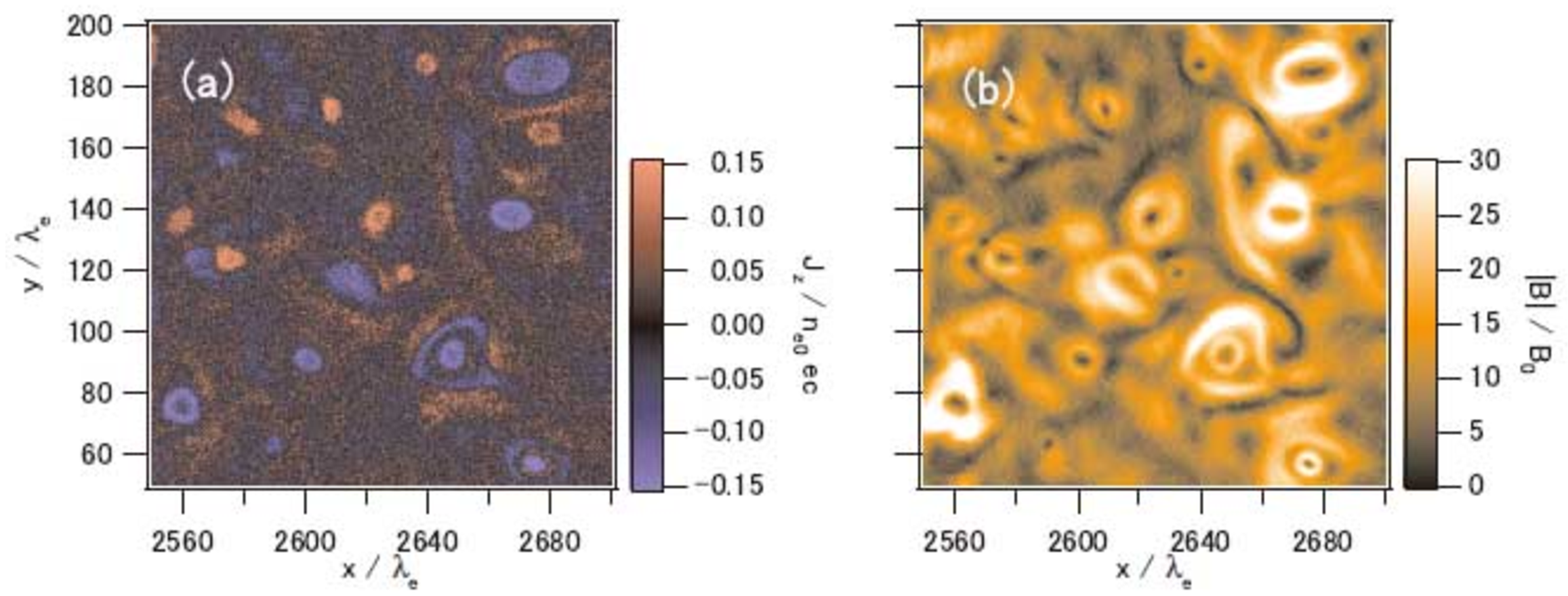}
\caption{
(a) Current density $J_z$ normalized by $\nez ec$
and (b) magnetic field strength $|B|$ normalized by the upstream field $B_0$
in the downstream region $2550 \lambdae < x < 2700 \lambdae$ and $50 \lambdae < y < 200 \lambdae$.
There are numerous current filaments carrying currents in the z-direction and
they mainly generate the downstream tangled magnetic field.
Some of the current filaments have a coaxial structure.
}
\label{fig:Jz_B_DS}
\end{figure*}

Figure \ref{fig:B_abs_map} shows the magnetic field strength
normalized by the upstream background magnetic field, $|B| / B_0$.
There are some strong, highly tangled magnetic fields
in the transition region ($|B|/B_0 \sim 40$) and the downstream region ($|B|/B_0 \sim 15$,
which is much larger than the magnetic field strength when merely compressed $\sim 4$).
Figure~\ref{fig:Jz_B_DS} shows enlargements of the current density in the $z$ direction, $J_z$,
and the magnetic field strength $|B|$ in a rectangular area in the downstream region
($2550 \lambdae < x < 2700 \lambdae$ and $50 \lambdae < y < 200 \lambdae$).
It is evident that the downstream tangled magnetic field is mainly generated by the current filaments in $J_z$.
The filaments have typical sizes of $\sim 2 - 4 \lambdai$,
where $\lambdai \equiv (\mi/\me)^{1/2} \lambdae$ is the ion inertial length,
which is slightly larger than the filament size in the foot region.
This can be explained by current filaments coalescing downstream
of the foot region.
Some of the current filaments have a complex coaxial structure in that
they are surrounded by return and anti-return currents.

Coalescence of current filaments that carry a current in the $x$-direction in the foot region
is inhibited due to the dimensionality of the simulation,
while current filaments in the $z$-direction
can merge with each other \citep[c.f.][]{Morse71, Lee73, Kato05}.
Furthermore, the current filaments in the $z$-direction are not affected by
instabilities in the current direction, such as the kink instability.
Therefore,
the current structures in the foot and downstream regions
may still differ in three dimensions.

\subsection{Early evolution}
\begin{figure}[htbp]
\plotone{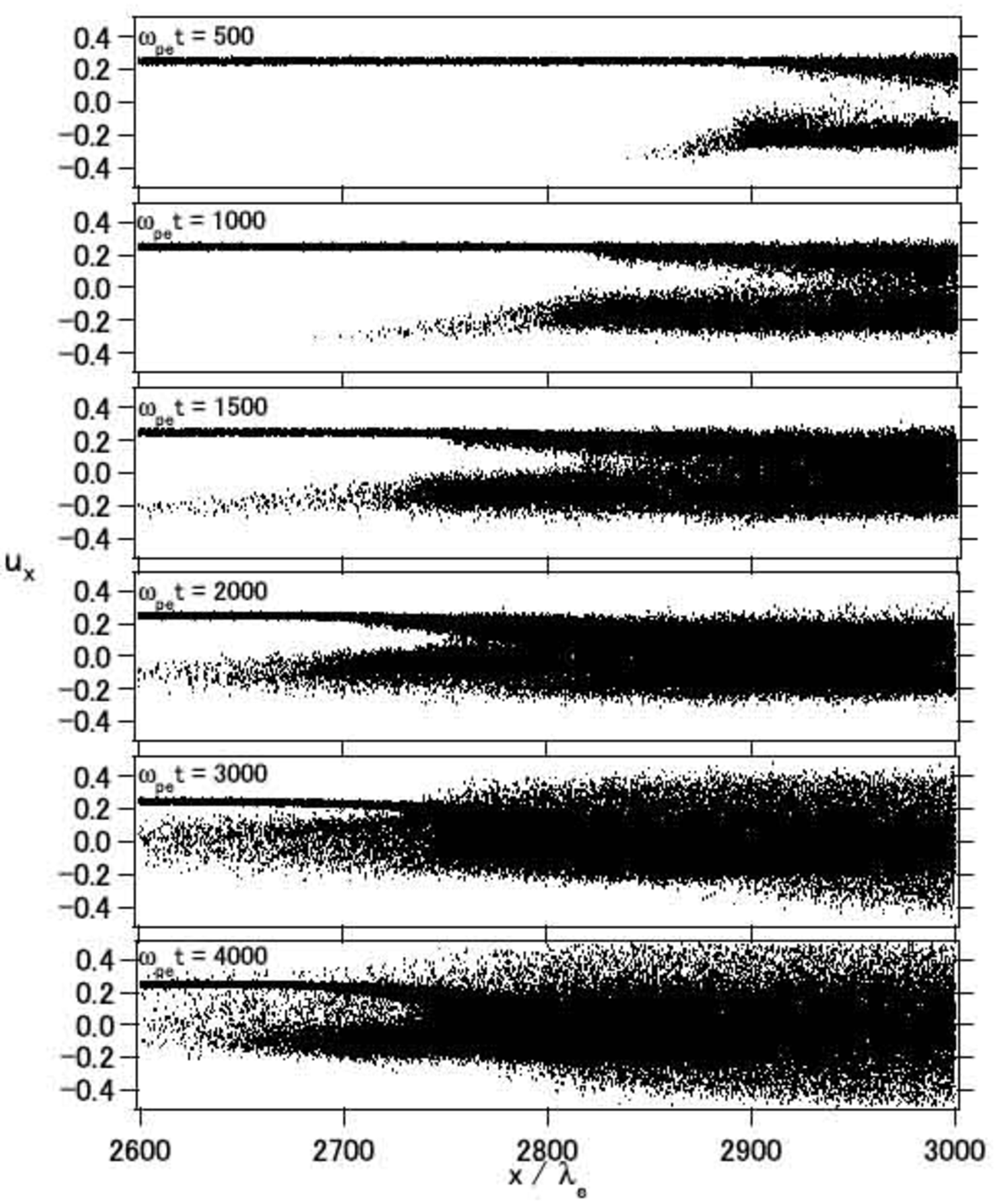}
\caption{
$x$--$u_x$ phase space plots of the ions in the early evolution.
Phase space plots at (from top to bottom) $\omegape t = 500$,
$1000$, $1500$, $2000$, $3000$, and $4000$.
} 
\label{fig:particles_early}
\end{figure}

\begin{figure}[htbp]
\plotone{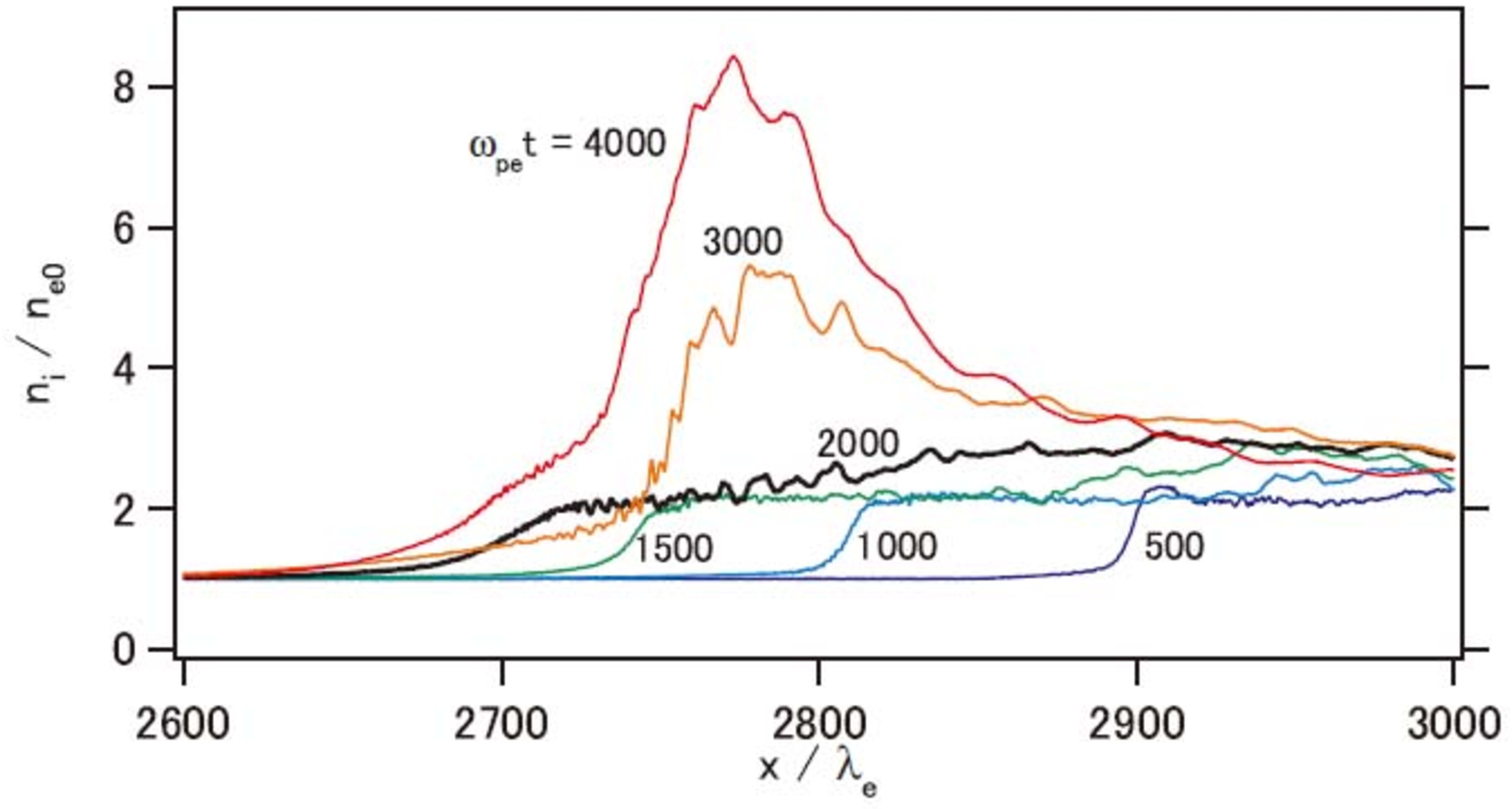}
\caption{
Time evolution of the ion number density
for $\omegape t = 500$,
$1000$, $1500$, $2000$, $3000$, and $4000$.
}
\label{fig:ni_prof_early}
\end{figure}

It is interesting to note the evolution of the system
before the effect of the background magnetic field becomes significant
(i.e., $t \le \Omega_i^{-1} \sim 2000 \omegape^{-1}$).
In this period, the plasma is effectively unmagnetized
and another kind of shock, namely an unmagnetized shock, appears.
Figures~\ref{fig:particles_early} and \ref{fig:ni_prof_early} respectively show
the evolution of the ions in the $x$--$u_x$ phase space
and that of the ion number density.
At a very early time ($\omegape t = 500$),
the incoming ions and the ions reflected by the wall at $x = 3000 \lambdae$
form a counterstreaming beam system and the number density in the overlapping region
simply becomes twice the upstream number density.
The two populations then start to interact ($1000 \le \omegape t \le 2000$)
and the density on the right side of the overlapping region increases.
As shown below,
this interaction is due to the magnetic field generated by
the ion beam--Weibel instability in the overlapping region
that deflects the ions; this provides a kind of dissipation mechanism.
At a later time ($3000 \le \omegape t \le 4000$),
the effect of the background magnetic field becomes important
and incoming ions start to accumulate around $x \sim 2775 \lambdae$.
Most of the ions that are initially reflected by the wall gyrate back downstream
due to the background magnetic field by $\omegape t = 3000$.
Instead,
at $\omegape t = 4000$, another reflected ion population appears around $x = 2700 \lambdae$; the ions in this population
have been newly reflected at the `ramp' ($x \sim 2750 \lambdae$).
Thus,
the structure changes from an unmagnetized shock to a magnetized one
around that time.

\begin{figure}[htbp]
\plotone{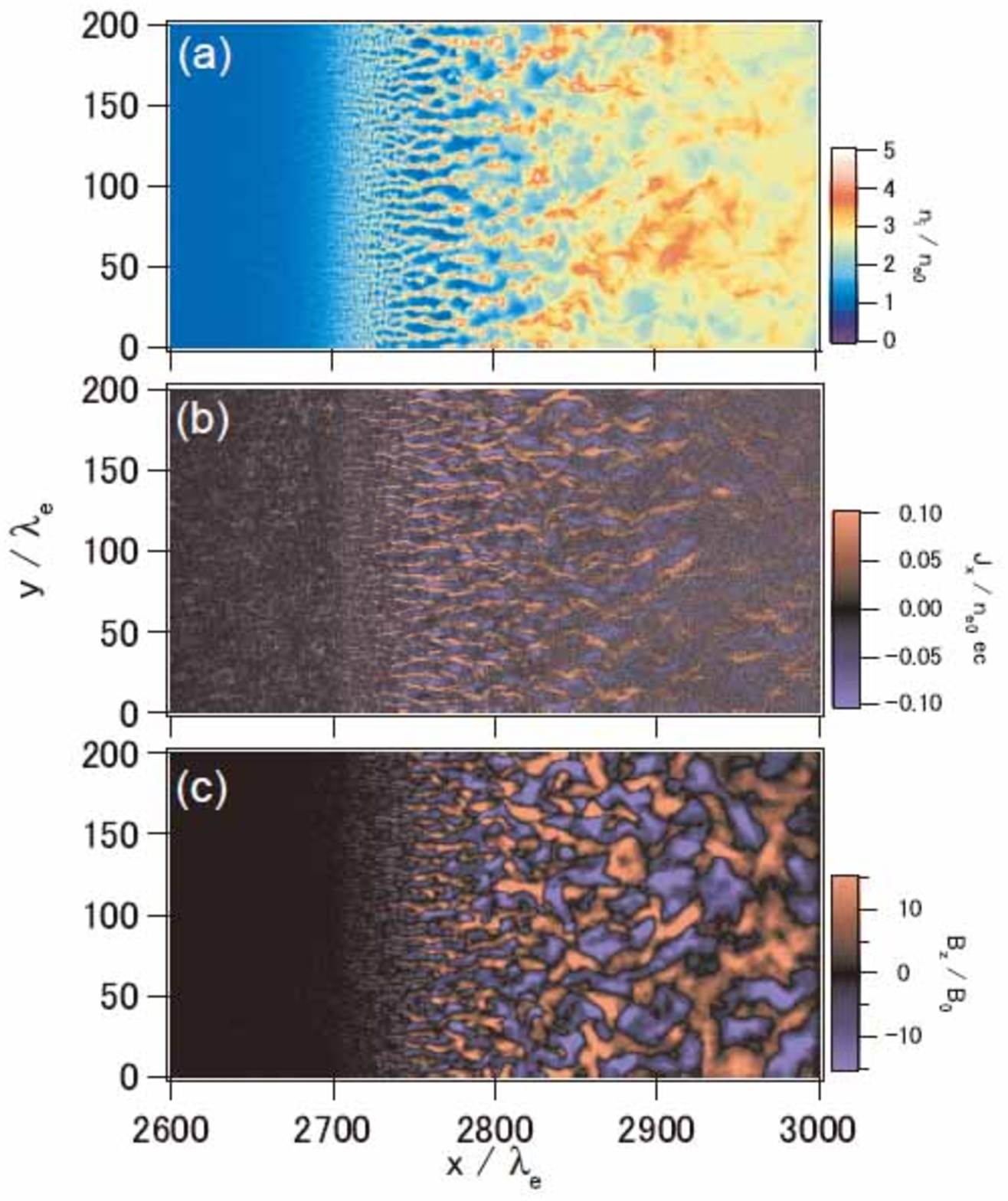}
\caption{
(a) The ion number density, (b) the $x$-component of the current density,
and (c) the $z$-component of the magnetic field
at $\omegape t = 2000$.
It is evident that many current filaments exist
and generate a strong magnetic field, which eventually
isotropizes the incoming ions. 
}
\label{fig:fields_t2000}
\end{figure}

\begin{figure}[htbp]
\plotone{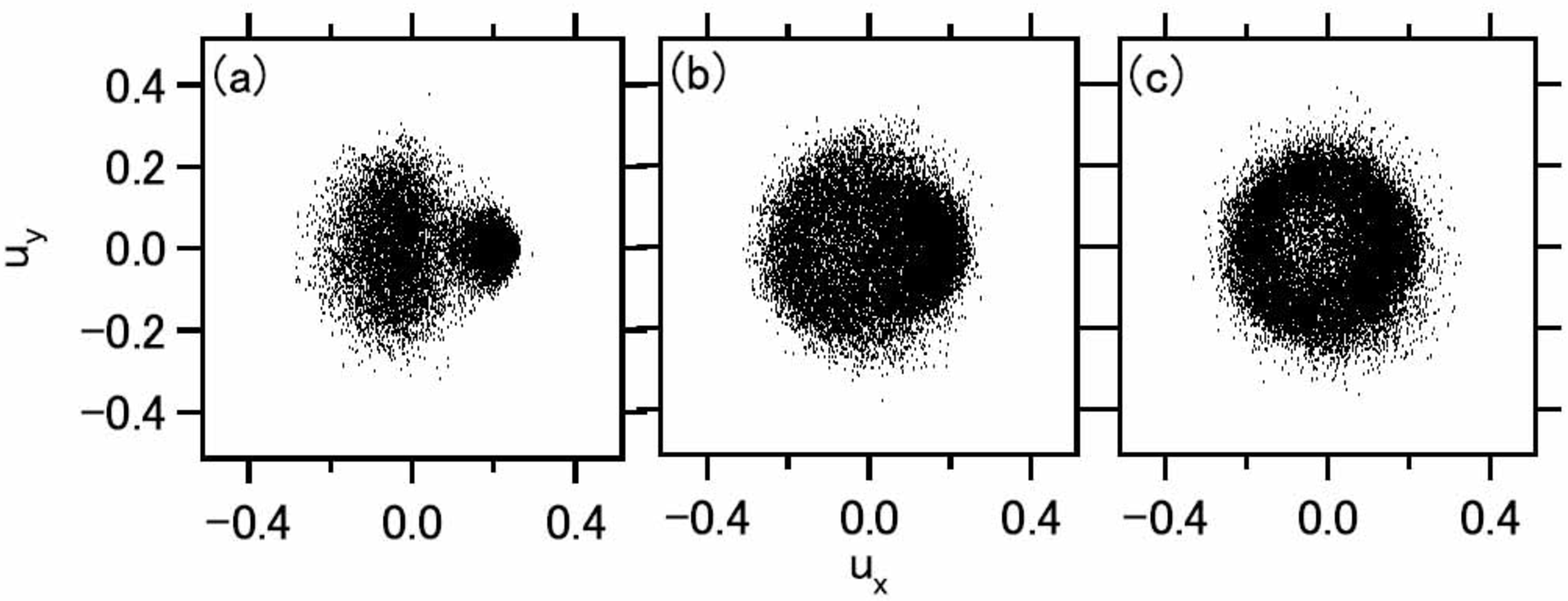}
\caption{$u_x$ -- $u_y$ plots of the ions at $\omegape t = 2000$ in three regions:
(a) $2700 < x/\lambdae < 2800$,
(b) $2800 < x/\lambdae < 2900$, and
(c) $2900 < x/\lambdae < 3000$.
In (c), the ions are almost isotropized
in the $u_x$ -- $u_y$ plane, forming a ring-like distribution.
}
\label{fig:ux_uy_t2000}
\end{figure}

Figure~\ref{fig:fields_t2000} shows
the ion number density, the $x$-component of the current density $J_x$,
and the $z$-component of the magnetic field $B_z$ at $\omegape t = 2000$.
It is evident that many current filaments are generated
in the interacting region and generate a strong magnetic field
around themselves.
A filamentary structure is also observed in the number density
corresponding to the current filaments.
These current filaments are generated by the ion beam-Weibel instability between
the counterstreaming ion populations.
The generated magnetic field
provides the dissipation mechanism for the ions to form the unmagnetized shock.
Figure~\ref{fig:ux_uy_t2000} shows $u_x$--$u_y$ plots in the following three regions:
$2700 < x/\lambdae < 2800$,
$2800 < x/\lambdae < 2900$,
and
$2900 < x/\lambdae < 3000$.
The ions are almost completely isotropized
and form a ring-like distribution in the $u_x$-$u_y$ plane
far downstream ($2900 < x/\lambdae < 3000$).
Since the beam-Weibel instability generates
only the $z$-component of the magnetic field in this 2D configuration,
the magnetic field deflects the ions only in the $u_x$-$u_y$ plane.
However, in three dimensions,
the generated magnetic field can have the $y$-component
and the ions can be deflected
in all directions, resulting in three-dimensional dissipation.

The dissipated ions form the downstream region of the shock
and the shock structure propagates upstream at an almost constant speed,
as shown in Fig.~\ref{fig:ni_dev} for $1500 \le \omegape t \le 2000$.
In the present simulation,
this Weibel-mediated shock disappears when the effect of the background magnetic field becomes significant at later times.
However,
this shock can propagate steadily upstream when there is no background field
\citep{Kato08}.

\section{DISCUSSION}

In the previous section it was found that
the filamentary structures in our simulation are generated by the ion beam--Weibel instability
in the foot region.
Similar structures have been found in the foot
or overshoot region
in 2D PIC or hybrid simulations with lower Mach numbers.
There are two different causes for these structures:
the emission of whistler waves at the ramp
\citep{Krauss-Varban95, Hellinger07, Lembege09}
and the emission of Alfv\'en waves 
due to the Alfv\'en ion cyclotron instability,
resulting in a structure called rippling
\citep{Winske88,Lowe03}.
Since these are associated with waves generated at the ramp or the overshoot and which 
then propagate upstream,
these processes could also be related to the filamentary structures
in our simulation.
The wavenumber range of the whistler wave is given by \citep{Ichimaru}:
\begin{equation}
	\frac{4\me}{\mi} \ll \frac{\nez}{n_\mathrm{e}} \left(\frac{kc}{\omegape}\right)^2 \ll 1,
	\label{eq:whistler_k}
\end{equation}
where $n_\mathrm{e}$ is the local electron number density
at which the whistler wave exists
and $\nez$ and $\omegape$ are defined for the far upstream.
The left-hand side is $\sim 0.133$ in our simulation
and if we take $n_\mathrm{e}/\nez \sim 2$ [Fig.~\ref{fig:V_prof}(c)]
and $kc/\omegape \sim 1$ [Fig.~\ref{fig:foot_filaments}(d)] as typical values,
the above condition is (marginally) satisfied.
However,
if the structure is related with (standing) whistler waves,
its group velocity must be greater than the shock speed (in the upstream frame):
\begin{equation}
	v_\mathrm{g}
    = \frac{\partial \omega}{\partial k}
	= 2 \frac{B}{B_0} \left(\frac{n_\mathrm{e}}{\nez}\right)^{-1} \frac{|\omega_\mathrm{ce}|}{\omegape} \frac{kc^2}{\omegape}
	> \Vsh,
\end{equation}
where $B_0$ and $\omegace$ are defined for the far upstream,
whereas $B$ is the local value.
[Here, we neglect the dependence of the propagation angle
and hence it is just a necessary condition;
c.f., \citet{Krauss-Varban95}.]
This condition can be rewritten as
\begin{equation}
	\MA < 2 \frac{B}{B_0} \left(\frac{n_\mathrm{e}}{\nez}\right)^{-1} \left(\frac{\mi}{\me}\right)^{1/2} \frac{kc}{\omegape}.
\end{equation}
In our simulation, the right-hand side is $\sim 10 - 30$, whereas the left-hand side is $\sim 130$.
Thus,
the filamentary structure observed in our simulation
does not originate from whistler wave emission.
On the other hand,
the result of the simulation by \citet{Lembege09} with $\mi/\me = 400$ satisfies this condition
if $B/B_0 \sim n_\mathrm{e}/\nez$;
the right-hand side is $\sim 2 kc/\omegapi \sim 14.6$, whereas the left-hand side is $\sim 4.93$.
Assuming $kc/\omegape \sim 1$ and $B \propto n_\mathrm{e}$,
a rough condition for the Alfv\'en Mach number is given by
\begin{equation}
	\MA < 2 (\mi/\me)^{1/2}.
\end{equation}
This gives
$\MA \le 11$ for the mass ratio in our simulation ($\mi/\me = 30$) and
$\MA \le 86$ for the real mass ratio of $\mi/\me = 1836$.
The rippling \citep{Winske88,Lowe03}
cannot be the cause of the filamentary structure in our simulation
because
(1) the rippling develops in the overshoot region,
whereas the filamentary structure in our simulation develops in the foot region;
(2) the wavelength observed in our simulation ($\lambda \sim 8\lambdae$)
is smaller than that of Alfv\'en waves
($\lambda_A > 3 \lambdai \sim 16 \lambdae$ for $\mi/\me = 30$);
and
(3) the Alfv\'en speed in the foot is smaller than the shock speed.
Nevertheless,
comparison of Fig.~\ref{fig:J_B} with the figures in \citet{Winske88}
reveals that
the structures of the magnetic field and the number density
at and behind the overshoot are similar to each other.
Therefore,
the rippling mechanism may also operate there in our simulation.
\citet{Lembege09} showed that
these structures can also be affected by the dimensionality of the simulations
in 2D simulations; the structures when the background magnetic field lies in
the simulation plane may differ from those when
it is perpendicular to the plane.

Electron acceleration was not observed in our simulation,
whereas \citet{Amano09} observed 
a kind of electron acceleration in a perpendicular shock
in their 2D PIC simulation, which used similar parameters to ours.
The main reason for this is considered to be the different directions of
the upstream background field:
in our simulation it lay in the simulation plane, whereas in the simulation by \citet{Amano09} it was out of the plane.
This again demonstrates that dimensionality can affect the results,
even for 2D simulations.

\begin{figure}[htbp]
\plotone{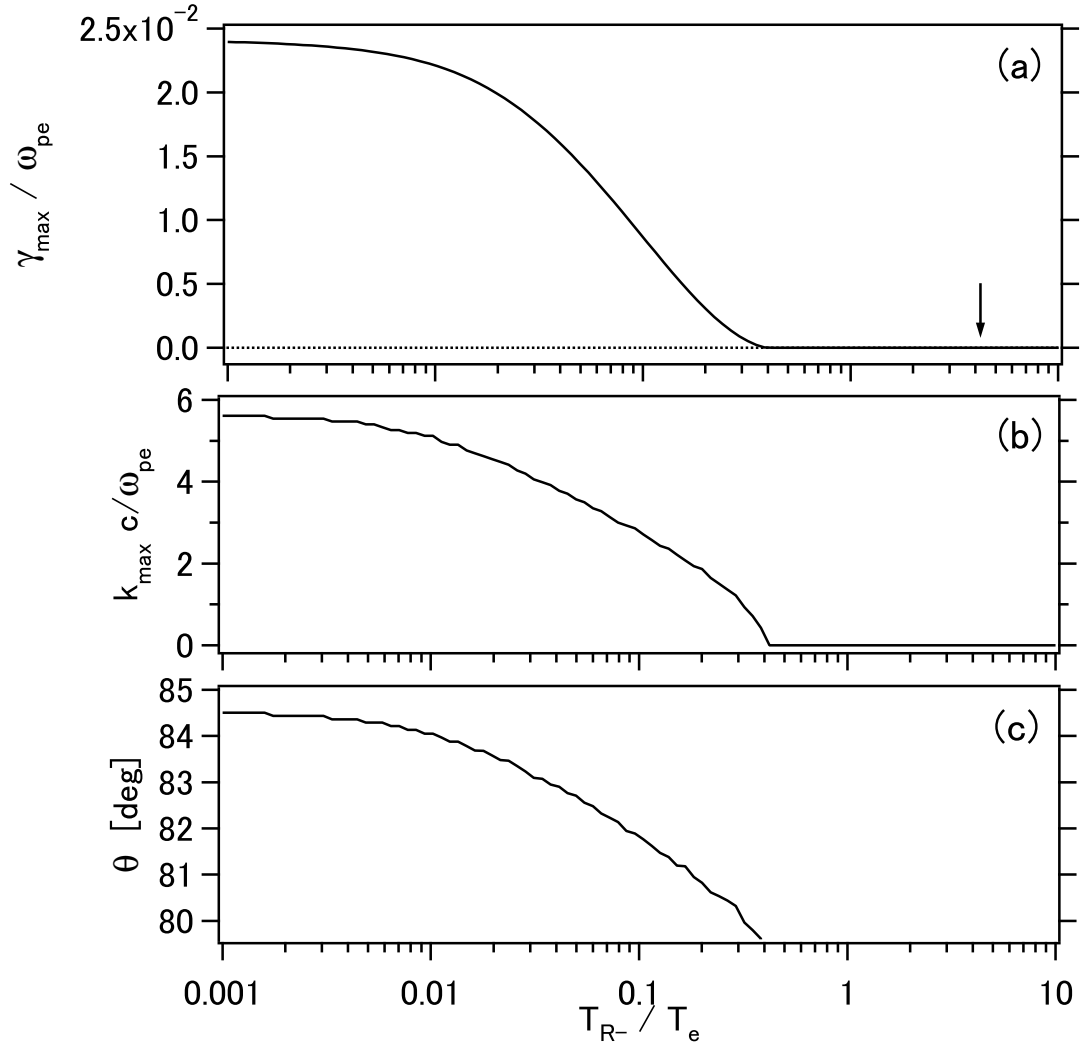}
\caption{
(a) The linear growth rate, (b) wavenumber, and (c) angle
of the most unstable mode of the ion--ion streaming instability
as functions of the temperature of the reflected ions streaming upstream, $T_\mathrm{R-}$,
with the other parameters obtained from the simulation at $\omegape t = 8000$ and $x = 2375 \lambdae$,
as in Fig.~\ref{fig:BI_x2300}.
The arrow indicates the initial value of $T_\mathrm{R-}$ in the simulation.
It is clear that the instability significantly depends on $T_\mathrm{R-}$.
}
\label{fig:IISI_x2375}
\end{figure}

The ion--ion streaming instability \citep{Stringer64, Ohnuma65, Forslund70},
which has a wavevector that is highly oblique to the streaming direction, 
can be driven by the interaction between the incoming ions and the reflected ions streaming upstream
in the foot region
and it can contribute to ion heating \citep{Auer71, Papadopoulos71, Wu84, Ohira08}.
[This instability can also be driven in front of electrostatic shocks
in two dimensions \citep{Kato10}.]
However,
this instability was not observed in our simulation (see the right panels in Fig.~\ref{fig:rho_power}).
This is due to the high temperature of the reflected ions streaming upstream, $T_\mathrm{R-}$,
in our simulation.
According to \citet{Ohira08},
the ion--ion streaming instability is efficient
for wavenumbers $k > \kDe$,
whereas it is damped for $k > \kDi$ 
due to the thermal motion of the ions.
Therefore, $\kDe < \kDi$ is a necessary condition for efficient growth
of the instability.
In the present case,
the two ion populations have different temperatures (Fig.~\ref{fig:T_prof})
and the Debye wavenumber of the reflected ions, $k_\mathrm{D,R_-}$, should be used to
evaluate the thermal damping effect.
Thus,
the condition for effective growth of the ion--ion streaming instability
is $\kDe < k_\mathrm{D,R_-}$ or
\begin{equation}
	\frac{T_\mathrm{R-}}{\Te} < \frac{n_\mathrm{R-}}{n_\mathrm{e}}.
\end{equation}
The right-hand side of this equation is always smaller than unity,
whereas the left-hand side is larger than unity in our simulation,
as shown in Fig.~\ref{fig:Vrel_profiles}.
For example,
at $x = 2375\lambdae$,
the ratios are $T_\mathrm{R-}/T_\mathrm{e} = 4.2$ and $n_\mathrm{R-}/n_\mathrm{e} = 0.32$,
respectively (see Table~\ref{table:sim_values}).
Hence,
the above condition is not satisfied
and this would explain why the instability does not grow in our simulation.
To examine the effect of $T_\mathrm{R-}$ more quantitatively,
we performed linear analysis with the parameters obtained from the simulation
at $\omegape t = 8000$ and $x = 2375 \lambdae$, but changing $T_\mathrm{R-}$
in the same way as in Fig.~\ref{fig:BI_x2300}.
Figure~\ref{fig:IISI_x2375} shows the linear growth rates,
the wavenumbers, and the angles between the wave vector and the streaming direction
of the most unstable mode
of the ion--ion streaming instability
as functions of $T_\mathrm{R-}$.
It shows that the instability depends strongly on $T_\mathrm{R-}$.
In particular, for $T_\mathrm{R-} / T_\mathrm{e} \ge 0.35$,
it does not grow at all.
Thus, the temperature of the reflected ions ($T_\mathrm{R-}$)
is important for growth of the instability in the foot region
as well as the Buneman instability.

Finally,
we mention the possibility of generating magnetized shocks in experiments.
Present large-scale laser facilities can generate
collisionless plasma flows at speeds of $\sim 1000$ km s$^{-1}$
\citep{Takabe08}.
Thus, if magnetized collisionless plasmas flowing at this velocity can be generated in laboratories,
it should be possible to perform experiments on magnetized shocks.
Table \ref{table:experiment} shows the required background magnetic field strengths $B_0$
for several sigma values and for a number density of $\nez = 10^{20}$ cm$^{-3}$
together with the corresponding ion gyration time, $T_g$, and ion gyro radius, $r_g$.
In this table,
the gyration time and the gyro radius are calculated for an ion mass of $1836 \me$
and a flow velocity of 1000 km s$^{-1}$.
Sigmas of $\tilde{\sigma} = 10^{-3}$ or $\tilde{\sigma} = 10^{-4}$
give achievable values for present large-scale laser facilities.
\begin{table}[!h]
\caption{
Magnetic field strength required for experiments
}
\begin{center}
\begin{tabular}{ccccc}
	\hline
	$\tilde{\sigma}$ & $\tilde{\MA}$ & $B_0$ (G) & $T_g$ (s) & $r_g$ (m)\\
	\hline
	$10^{-2}$ & $10$ & $4.6 \times 10^5$ & $1.4 \times 10^{-9}$ & $2.3 \times 10^{-4}$\\
	$10^{-3}$ & $32$ & $1.4 \times 10^5$ & $4.7 \times 10^{-9}$ & $7.3 \times 10^{-4}$\\
	$10^{-4}$ & $100$ & $4.6 \times 10^4$ & $1.4 \times 10^{-8}$ & $2.3 \times 10^{-3}$\\
	$10^{-5}$ & $320$ & $1.4 \times 10^4$ & $4.7 \times 10^{-8}$ & $7.3 \times 10^{-3}$\\
	\hline
\end{tabular}
\end{center}
\label{table:experiment}
\end{table}

\section{CONCLUSION}
We performed a 2D PIC simulation
to investigate collisionless shocks propagating in weakly magnetized
electron--ion plasmas at nonrelativistic speeds
with a sigma of $\tilde{\sigma} = 10^{-4}$.
We showed that current filaments are generated within
the foot region by the ion beam--Weibel instability
and that they generate
magnetic fields in the same manner as for Weibel-mediated shocks in unmagnetized plasmas.
The magnetic field strength generated by the current filaments
is comparable with or even stronger than those of the compressed background magnetic
field. Therefore, these filaments and their associated magnetic fields,
which cannot be analyzed by 1D simulations,
are important in the formation of collisionless shocks in weakly magnetized plasmas.
There are current filaments in the downstream region and they generate a tangled
magnetic field that is typically 15 times stronger than the upstream background field.
The thermal energies of electrons and ions in the downstream region
are not in equipartition and their temperature ratio is given by $\Te/\Ti \sim 0.33$.
We found a fraction of the ions were slightly accelerated on reflection at the shock,
whereas significant electron acceleration was not observed in our simulation.
The simulation results agree very well with the Rankine--Hugoniot relations.
It was also shown that electrons and ions are heated in the foot region
by the Buneman instability (for electrons) and the ion-acoustic instability
(for both electrons and ions).
However, the growth rate of the Buneman instability was significantly reduced
from typical growth rates of this instability
because of
the relatively high temperature of the reflected ions.
For the same reason,
ion--ion streaming instability did not grow in the foot region.

\acknowledgments
One of the authors (T.N.K.) is grateful to A. Spitkovsky for helpful discussions.
We also thank Y. Sakawa for discussions about their experiments.
This work was supported in part by the Ministry of Education, Culture, Sports, Science and Technology (MEXT), Grant-in-Aid for Young Scientists (B) (T.N.K.: 20740136 and 22740164),
and in part by the National Science Foundation (Grant No. NSF PHY05-51164).
Numerical computations were performed at the Cybermedia Center, Osaka University, Japan.


\end{document}